\documentclass[preprint,authoryear,11pt]{elsarticle}
\usepackage[letterpaper,left=1in,right=1in,top=1.35in,bottom=1.5in]{geometry}
\usepackage{amsmath,amssymb}
\usepackage{graphicx}
\usepackage{multirow}
\usepackage{booktabs}
\usepackage{xcolor}
\usepackage[hidelinks]{hyperref}

\usepackage{xr-hyper}
\usepackage{algorithm}
\usepackage{float}
\usepackage{algpseudocode}

\journal{Expert Systems With Applications}

\begin{document}

\begin{frontmatter}

\title{Risk-Aware Generative Inpainting for Optimized Design Editing of EV Battery Cooling
Channels}

\author[kaist]{Leekyo Jeong}
\ead{leekyoj@kaist.ac.kr}
\author[kaist]{Yoon Koo Lee}
\author[kaist,narnia]{Namwoo Kang\corref{cor1}}
\ead{nwkang@kaist.ac.kr}
\cortext[cor1]{Corresponding author.}
\address[kaist]{Cho Chun Shik Graduate School of Mobility, Korea Advanced Institute of Science
and Technology (KAIST), Daejeon 34051, Republic of Korea}
\address[narnia]{Narnia Labs, Daejeon 34051, Republic of Korea}

\begin{abstract}
Cooling-channel layouts for electric-vehicle battery packs must deliver temperature uniformity
and low pressure drop while maintaining a single continuous channel. In late-stage design,
local topology modification offers a practical way to improve performance while retaining
established global features. Diffusion-based inpainting naturally accommodates such local
edits, but its stochastic nature can produce different outcomes even when applied to the same
region. This variability poses a fundamental question: how should edit locations be selected
when the outcome of each modification is stochastic? We propose a risk-aware generative
editing framework that incorporates this variability into the selection of edit locations.
Instead of scoring each candidate location by a single expected improvement, the method
estimates a distribution of possible outcomes from offline edit results evaluated using a
computational-fluid-dynamics (CFD)-trained surrogate and ranks locations according to a chosen
level of risk. A single
trained model therefore supports different editing preferences at inference time, emphasizing
either higher expected improvement or greater consistency without retraining. The policy is
evaluated against random editing in a held-out paired study across seven mask configurations.
It improves the cooling-channel objective over random editing in most configurations, and the
advantage is retained under independent CFD verification of the edited designs. Varying the risk level reveals a consistent trade-off between mean improvement
and run-to-run consistency, while the learned distribution is useful for ranking locations but
should not be interpreted as a calibrated probability distribution. Together, these results show that stochastic generative editing can be converted from a source
of variability into a controllable design decision through risk-aware location selection:
effective editing depends not only on how a design is modified, but also on where it is
modified and how much outcome variability is acceptable.
\end{abstract}


\end{frontmatter}


\section{Introduction}
\label{sec:intro}

Many engineering design problems can be formulated as iterative refinement processes in which an
existing feasible solution is progressively modified to improve its performance. Rather than
replacing the entire design at each iteration, selected regions can be modified while satisfactory
portions of the current solution are retained. This formulation introduces a decision problem at
each iteration: which region should be modified, and what modification should be considered at
that location?

The problem becomes more challenging when a given location admits multiple plausible
modifications. A single region may admit alternative modifications with substantially different
design outcomes. Consequently, selecting the next edit requires comparing candidate locations
while accounting for the range of outcomes that their possible modifications may produce.
Generative models provide a natural mechanism for exploring such alternatives
\citep{regenwetter_review}, since a local region can be regenerated into multiple plausible
configurations while the remainder of the design is preserved.

Electric-vehicle (EV) battery cooling-channel design provides a representative engineering
setting for this problem. Thermal management is a central constraint in the design of EV battery
packs and high-power electronics --- ambient and operating-temperature extremes alone can shift
an EV's usable driving range by tens of percent \citep{yuksel2015regional} --- and embedded
cooling channels have to provide low and \emph{uniform} component temperatures, small pressure
drop, and a manufacturable layout at once. Cooling-channel layouts are discrete connected
topologies, and improving an existing layout can be formulated as a sequence of local
modifications: a region is selected, edited, and evaluated while the remainder of the layout is
retained. Each such edit must honour a \emph{single-stroke connectivity} constraint (one
continuous channel, a manufacturing requirement) that classical adjoint topology optimization and
parametric optimization both handle poorly on a discrete pixel grid. Importantly, these
constraints define the feasible engineering design space rather than merely restricting the
optimization problem. Accordingly, the key challenge is not to remove these constraints, but to
efficiently identify promising edit locations within the feasible space.

Diffusion-based inpainting instantiates this local-editing mechanism directly: masking a
selected region and regenerating its contents can modify part of an existing channel while
preserving the surrounding design, and diffusion-based generators in particular
\citep{ho2020ddpm} can synthesize completions that already satisfy the single-stroke topology.
These generators are \emph{stochastic}, however: the same edit mask produces different
completions on every run. Two questions follow at once: (i) which edit sites deserve an
inpainting call in the first place, and (ii) how should an agent cope with that per-call
stochasticity?

Both questions are answered here with the same mechanism: a distributional value function that
scores each candidate edit site by its outcome distribution, not its expected value alone.
This is the outlook of distributional reinforcement learning (RL), which supports risk-sensitive
decisions that act on the tails of the distribution instead of its mean alone (reviewed in
\S\ref{sec:rw_dist}). Within that family of architectures we adopt the implicit quantile network
(IQN), which can be queried at an arbitrary probability $\tau$ without committing to any fixed
set of quantile fractions \citep{iqn2018}, so one trained network exposes a whole family of
risk-sensitive policies through the value of $\tau$ alone.

IQN is conventionally trained end-to-end by sequential reinforcement learning:
temporal-difference (TD) learning with a bootstrapped target, $n$-step returns, and a replay
buffer. For this editing task we found that the paradigm degrades the learned distribution
(\S\ref{sec:tdvsup}), so we dropped it while keeping IQN's query interface.

In this study, diffusion inpainting serves as the edit primitive of a design-editing agent for
cooling-channel skeletons. A frozen ridge-regression term on five geometric path features is
combined with a $\tau$-conditioned residual network, both fit by \emph{supervised} regression on
offline-labelled edit outcomes with no bootstrapping or replay; at inference, every candidate
edit site is scored once and visited in that fixed order for a fixed number of steps, a
deterministic single-pass policy, with $\tau$ the inference-time risk dial detailed in the
contributions below. The policy beats random editing across most mask configurations under a
pre-registered evaluation protocol, an advantage that survives independent CFD verification and
is not explained by surrogate scoring error, at a measurable cost in seed-to-seed consistency and
in how faithfully the learned distribution calibrates as a probability object. Comparisons of
this kind also face two confounds invisible to ordinary testing --- an input-convention mismatch
that silently decorrelates the training objective from ground truth, and a rollout-seed
dependence that leaves any single-realization number an incomplete summary of the effect
(\S\ref{sec:rolloutseed}) --- both measured and reported throughout this study.

Existing approaches primarily optimize or generate designs, whereas our framework addresses
the decision of where to perform a stochastic local generative edit on an already feasible
topology. This study makes four contributions:
\begin{enumerate}
  \item A single inference-time risk dial $\tau$ that moves the deployed policy along a
        mean-improvement/seed-to-seed-consistency frontier without retraining. A mixed-effects
        decomposition locates the measurable benefit in this dial: the effect is robust across a
        three-seed IQN-vs-FQF comparison, and provisional for the single-seed DQN and ridge-only
        arms (\S\ref{app:arch}).
  \item \label{contrib:c1} A held-out, paired evaluation of the trained policy against random
        editing across seven mask configurations, with effect sizes reported both at a
        pre-registered rollout seed and averaged over independent rollout seeds. We evaluate
        against random editing only; comparison against a published cooling-channel RL agent
        \citep{kimhan2023} was judged out of scope because its action space and training code are
        not directly transferable (\S\ref{sec:rw_cooling}).
  \item A CFD-verified fidelity study of the surrogate objective that drives training: high
        decision-level agreement with CFD within the deployment range, and no statistical
        association between the measured policy advantage (Contribution~\ref{contrib:c1}) and
        surrogate scoring error, evidence against reward exploitation as an explanation for that
        advantage.
  \item A methodological account, null results included, of what the risk-conditioned quantile
        function can and cannot do: a temporal-difference-trained version is measurably
        worse-calibrated than the same architecture trained by supervised regression, and even the
        supervised version, useful for \emph{ranking} edit sites, fails a standard calibration
        test structurally and must not be read as a calibrated probability distribution.
\end{enumerate}

Because diffusion sampling introduces substantial rollout-level stochasticity, we explicitly
evaluate the policy across independent rollout seeds rather than relying on a single realization.

The rest of the paper runs as follows. Section~\ref{sec:related} reviews related
work in cooling-channel design optimization, generative models for engineering design, and
distributional reinforcement learning. Section~\ref{sec:method} presents the proposed three-stage
system, first as a whole and then stage by stage: the objective, the surrogate evaluator, the
inpainting edit operator, and the learned position-value policy. Section~\ref{sec:setup} describes
the experimental setup: data generation, the surrogate model, and the evaluation
protocol. Section~\ref{sec:results} presents the measurements stage by stage, from surrogate fidelity
(Stage 1) and what iterative inpainting search achieves without a learned policy (Stage 2) to
the learned policy itself (Stage 3), covering held-out improvement and its rollout-seed
dependence, a factor decomposition of where the benefit lives, CFD verification on edited
designs, the comparison against TD training, and the calibration tests.
Section~\ref{sec:discussion} interprets these findings and states the limitations, and
Section~\ref{sec:conclusion} concludes.

\section{Related Work}
\label{sec:related}

\subsection{Cooling-channel and flow-field design optimization}
\label{sec:rw_cooling}
Cooling-channel and flow-field layout is a long-standing engineering-design problem; in
EV applications it is driven by the need to hold battery-pack temperatures
low and uniform while containing the pumping pressure drop. Existing methods broadly fall into
three categories: parametric optimization of a fixed channel template, typically a serpentine
cooling plate, against a thermal-hydraulic objective
\citep{jarrett2011,jarrett2014,yu2009serpentine,patil2023,benabdelaziz2020}; topology
optimization (TO), which lifts the template restriction by making the material distribution
itself the design variable
\citep{leeTO,yaji2018flowfield,chen2019vrfb,yajiTO}; and surrogate-assisted global search, which
couples a cheap learned model with evolutionary or generator-based search over candidate layouts
\citep{leafdesign2024,wan2022flowfield,zhong2025wgantl,srivardhan2026bioinspired,
ebbspicken2024deepedh,liu2026digitaltwin,he2023surrogate}. All three work well inside a predefined
parameterization but are less suited to free-form editing of a discrete single-stroke channel
skeleton: adjoint-based TO targets continuous density fields, not the \emph{binary} pixel grid and
hard single-stroke manufacturing constraint this problem imposes, and the surrogate-assisted line
still optimizes, screens, or generates inside a fixed structural parameterization rather than
making agent-directed local edits on an arbitrary skeleton.

The closest precedent is the reinforcement-learning cooling-channel agent of
\citet{kimhan2023}, in which a deep Q-network (DQN) grows a channel layout on a
grid step by step, calling CFD directly inside the training loop. We score each edit candidate
cheaply with a
deep-learning surrogate (\S\ref{sec:surrogate}), which is what keeps the editing loop affordable at
training scale, and our final policy, unlike that work, is not trained by sequential RL at
all (\S\ref{sec:mechanism}). Two further differences separate it from ours: its action space is a
hand-crafted grid move (no generative model), and its critic is a \emph{scalar} value function
blind to the return distribution and to risk. We do not attempt a quantitative head-to-head
comparison against this method: its action space and training code are not directly transferable
to our diffusion-based editing setting.

\subsection{Generative models for engineering design and diffusion inpainting}
\label{sec:rw_generative}
Deep generative models are now a standard tool for engineering design \citep{regenwetter_review}:
an early, influential example couples a generative adversarial network with topology optimization
to synthesize structural designs directly \citep{oh2019deepgen}, and the idea has since spread to
building-scale layout synthesis from fused text/image conditioning \citep{liao2022gan} and to
physics-constrained settings that couple a generator with physical-parameter generators through a
physics-based loss \citep{yu2024pggan}. In each case the generator proposes and a physical or
optimization-derived criterion disciplines what it proposes, a division of labour our system
shares as well.
Among the generative families, denoising diffusion probabilistic
models (DDPMs) \citep{ho2020ddpm} learn to reverse a gradual noising process and have become the
dominant image generator; their sampling is inherently \emph{stochastic}, a property that is
central to our formulation. Most relevant here is diffusion-based inpainting: RePaint
\citep{repaint2022} completes a masked region without retraining, repeatedly resampling the masked
region during the reverse process while keeping the known region fixed, which yields diverse
completions for an arbitrary mask. Our edit primitive descends directly from this idea:
``place a mask and let an unconditional DDPM regenerate that region.'' Related restoration-as-
inpainting formulations, such as the denoising diffusion null-space model (DDNM)
\citep{ddnm2023}, enforce data consistency through a null-space projection.

Diffusion models have also served directly as design generators, in structural topology
optimization, thermal heat-sink synthesis, and cold-plate digital twins
\citep{topodiff2023,heatgen2026,wu2026coldplate}: all cast design generation as a single,
conditional, surrogate- or gradient-guided sampling pass, so a design emerges from one guided
generation. We instead invoke an \emph{unconditional} generator repeatedly at agent-selected
sites and treat the resulting generative noise as a random quantity to be actively ranked and
risk-managed, not steered away by a guidance term.

A complementary line fine-tunes the generator's own weights toward a reward with RL, reading the
denoising chain as a multi-step decision process: DDPO \citep{ddpo2024} and
DPOK \citep{dpok2023}. Our use of the pretrained generator is orthogonal to that line --- we hold
it fixed and learn instead where, and at what scale, to apply it --- and the outer decision
function is not trained by RL either, for the reasons argued in \S\ref{sec:whynotrl}.

\subsection{Distributional reinforcement learning and risk-sensitive control}
\label{sec:rw_dist}
Standard value-based RL (e.g.\ DQN \citep{mnih2013dqn}) learns only the \emph{expectation} of the return,
$Q^{\pi}(s,a) = \mathbb{E}[Z^{\pi}(s,a)]$. Distributional RL models the full distribution
of the return $Z^{\pi}(s,a)$ instead, which opens the way to \emph{risk-sensitive} policies that act on the tails
of the distribution (e.g.\ CVaR), not merely its mean. \textbf{C51} \citep{c51} represents the
return distribution as a categorical distribution over 51 fixed atoms; \textbf{QR-DQN}
\citep{qrdqn} fixes a uniform set of quantile fractions and regresses the corresponding
quantile locations. \textbf{IQN} \citep{iqn2018} goes further by committing to no fixed
set of quantile fractions: a fraction $\tau\sim U(0,1)$ is sampled, embedded (cosine/Fourier
embedding), and fed into the network together with the state, so a single implicit network
approximates the whole quantile function and can be queried at arbitrary resolution.
\textbf{FQF} \citep{fqf2019} extends IQN by also \emph{learning} the quantile fractions via a
separate fraction-proposal network.

\emph{What we borrow from this literature, and what we leave behind.} We keep IQN's core interface
--- condition a network on a cosine-embedded $\tau$ and read off a value at that quantile --- as
the mechanism behind a single, re-trainable risk dial, but not the training paradigm it is
usually paired with (temporal-difference learning, bootstrapped targets, replay); \S\ref{sec:whynotrl} argues why.

On top of these distributions we place the conditional value-at-risk (CVaR),
introduced by \citet{rockafellar2000cvar} and belonging to the family of
coherent risk measures \citep{artzner1999}. Tail-risk criteria have been brought into sequential decision
making by \citet{tamar2015cvar}, \citet{chow2018cvarrl},
and \citet{keramati2020}, and in continuous control by distributional actor--critic
methods \citep{dsac2020}. Applied work in the same spirit builds risk or safety directly into the
decision rule of an industrial agent, as in the risk-sensitive controller for human--robot
collaborative manufacturing of \citet{wang2026risksensitive} and the safety-thresholded policy
selector of \citet{pan2026saferl}. Our deployment differs from each of these: the
risk-conditioned value function never passes through an RL training loop, because $\tau$ is a
supervised-regression conditioning input consulted only at inference to select a risk band. The
supervised half of that construction has its own history: \citet{xu2017cqrnn} fit
several quantiles jointly inside one network, which is the estimation problem our residual solves,
with the difference that our quantile fraction is sampled and embedded instead of fixed in
advance.

\section{Proposed System}
\label{sec:method}

\subsection{Overview of the proposed system}
\label{sec:system}
\begin{figure}[htbp]\centering
\includegraphics[width=\textwidth]{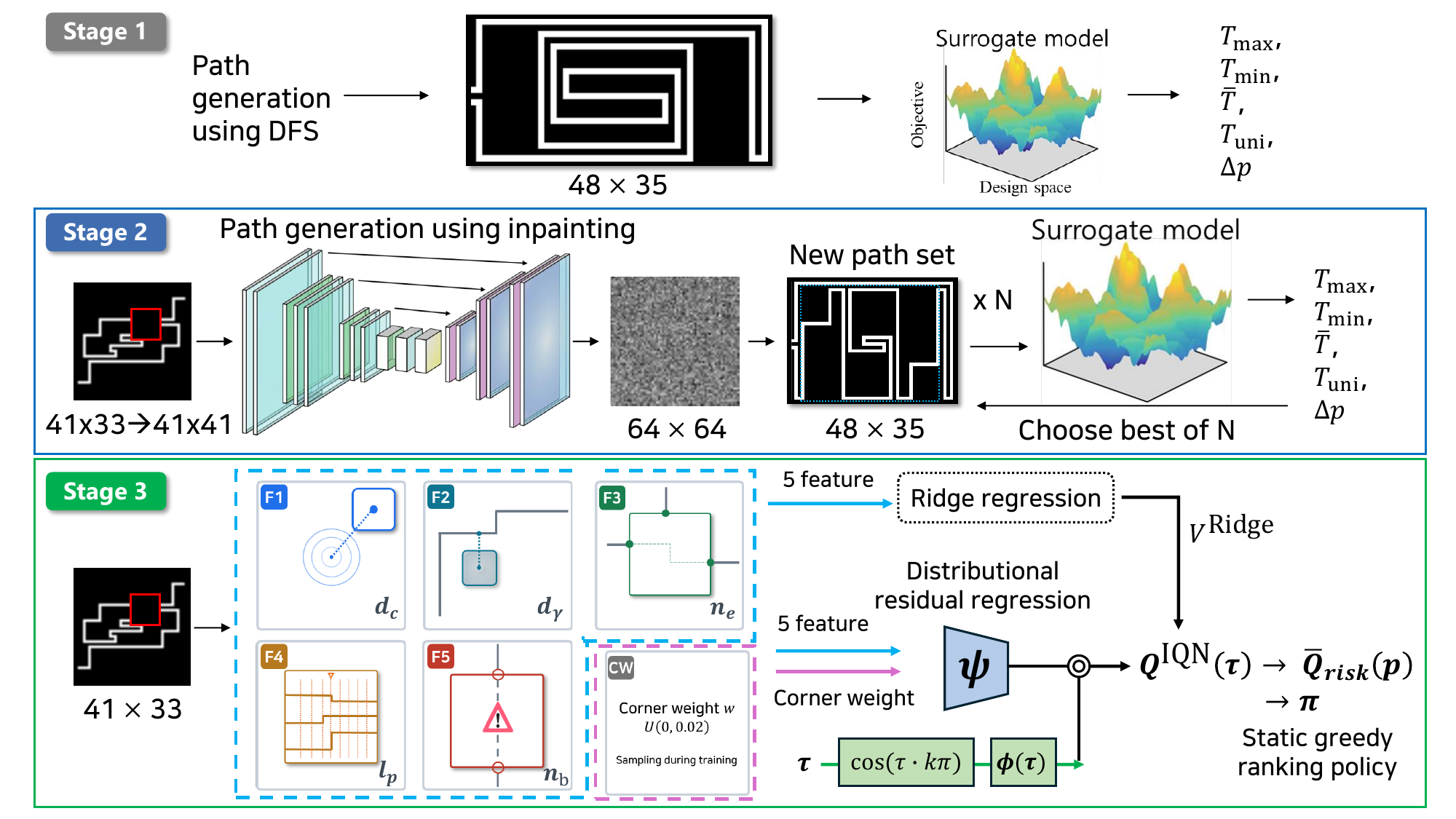}
\caption{Proposed cooling-channel design-support system: the current skeleton is edited by a
DDPM inpainting operator at a position selected by the static greedy ranking policy, the edited
layout is scored by the surrogate, and the policy's priority list is built from the frozen ridge
term plus the $\tau$-conditioned residual; CFD closes the loop offline as the verification stage.}
\label{fig:pipeline}
\end{figure}
The system is a three-stage pipeline, and these stage labels are used throughout the results.
Each stage answers the question the next one needs answered: Stage 1 settles how a layout is
evaluated, Stage 2 settles whether editing can improve it at all, and Stage 3 settles where to
edit first when the editing budget is small. Architecture and training details of the diffusion
inpainting operator itself are given in the supplementary material (\S\ref{app:inpaint}).
\textbf{Stage 1 (data generation and surrogate)}: layouts are generated, labelled by CFD, and a
surrogate is trained to predict performance for any layout in real time. \textbf{Stage 2
(iterative optimization by inpainting)}: a DDPM inpainting operator, driven by a best-of-$N$
search over randomly placed masks, iteratively improves a single design (Algorithm~\ref{alg:stage2}).
No learned position policy is involved here, and this stage also supplies the offline candidate
outcomes that \S\ref{sec:agent}'s position-value model is fit on. \textbf{Stage 3 (learned editing
policy)}: given a masked region's geometric features, the ridge $+$ residual model ranks candidate
edit positions by expected improvement under a chosen risk level, and the static policy visits them
in that order (Algorithm~\ref{alg:stage3}). Stage 3 is the contribution of this paper; Stages 1 and
2 come first because they establish the two premises it rests on.

The remainder of this section follows the pipeline in order: \S\ref{sec:score} fixes the design
representation and the objective, \S\ref{sec:surrogate} the Stage 1 evaluator,
\S\ref{sec:stage2method} the Stage 2 edit operator and search, and
\S\ref{sec:agent}--\S\ref{sec:policy} the Stage 3 position-value model, its training rule, and the
deployment policy.

\subsection{Design problem and objective score}
\label{sec:score}
The design is a binary single-stroke channel skeleton on a discrete grid, one pixel of which is
$10\times10\,$mm. The skeleton the policy edits occupies a $41\times33$ grid and carries no
inlet/outlet routing; that routing is attached afterwards by the assembly step
(\S\ref{sec:data}), giving the $48\times35$ domain that is actually scored and meshed, i.e.\ a
$480\times350\,$mm cold plate. The objective is a
composite, lower-is-better fitness score
\[
  \text{Score} \;=\; T_{\text{avg}} + T_{\text{uni}} + \Delta p + w_{\text{corner}}\cdot(\text{corners}),
\]
with all terms normalized and $w_{\text{corner}}=0.001$ at deployment (the corner term penalizes
manufacturability); \emph{improvement} $=$ initial score $-$ final score. $T_{\text{avg}}$ is the
volume-average metal temperature, $T_{\text{uni}}=2\,T_{\text{avg}}(1-\mathrm{UI})$ is the
top-surface temperature non-uniformity, and $\Delta p$ is the pressure drop across the channel.

\subsection{Stage 1: surrogate evaluator}
\label{sec:surrogate}
An EfficientNetV2-S surrogate, trained on CFD labels (\S\ref{sec:data}), scores each edited
candidate in place of CFD inside the scoring loop; CFD is kept for offline verification
(\S\ref{sec:cfd_verify}). Without this substitution the loop would be unaffordable: one
conjugate-heat-transfer case takes on the order of ten minutes of wall-clock time on 64 cores,
against milliseconds for a surrogate forward pass. A single 20-step rollout scores 20 candidates,
already more than three hours of 64-core time if each were sent to CFD, and the
evaluation campaign behind \S\ref{sec:results} scores on the order of $10^{6}$ candidates in
total. CFD is therefore reserved for the places where it is decisive, generating the training
labels
(\S\ref{sec:data}) and verifying final designs (\S\ref{sec:cfd_verify}), and never appears inside the
edit loop.

\subsection{Stage 2: inpainting edit operator and best-of-$N$ search}
\label{sec:stage2method}
The edit primitive shared by Stages 2 and 3 is an unconditional denoising diffusion model
applied as inpainting. The $41\times33$ skeleton is padded to $41\times41$ and resized to the
$64\times64$ working canvas; the masked window is regenerated by a 10-step reverse diffusion
schedule while the known region is held fixed, and the result is binarised and mapped back to
the skeleton frame, after which the inlet/outlet routing is attached to form the $48\times35$
layout the surrogate scores. A candidate is accepted only if the edited skeleton is still a
single continuous stroke with valid inlet/outlet attachment; up to five sampling attempts are
made, and a step in which no attempt is valid leaves the layout unchanged. Architecture and
training details of the operator are given in the supplementary material
(\S\ref{app:inpaint}).

Stage 2 uses this operator with no learned position policy: at each generation a mask is placed
at random, $N{=}8$ candidates are generated and scored by the surrogate, and the best-so-far
design is kept (Algorithm~\ref{alg:stage2}). Stage 2 serves two purposes. It establishes that
inpainting edits can improve a layout at all (\S\ref{sec:stage2}), and its per-position
candidate outcomes are the offline labels on which the Stage 3 position-value model is fit
(\S\ref{sec:trainconfig}).

\begin{algorithm}[H]
\caption{Stage 2 --- best-of-$N$ optimization by inpainting}
\label{alg:stage2}
\begin{algorithmic}[1]
\Require initial bare skeleton $x^{(0)}$ ($41{\times}33$); generations $G{=}200$; candidates/gen $N{=}8$; corner weight $w_{\text{corner}}$
\For{$g = 1, \dots, G$}
  \State sample a mask position $m_g$ uniformly at random over $x^{(g-1)}$
  \For{$i = 1, \dots, N$}
    \State $c_i \gets$ DDPMInpaint$(x^{(g-1)}, m_g)$ \Comment{$41{\times}33{\to}41{\times}41{\to}64{\times}64$, 10-step, $\le\!5$ attempts (\S\ref{app:inpaint}); $c_i{\gets}x^{(g-1)}$ if none valid}
    \State attach Type-B routing to $c_i$ ($\to 48{\times}35$); $s_i \gets \mathrm{Score}(c_i)$ (Eq.~\eqref{eq:qmodel}, weight $w_{\text{corner}}$)
  \EndFor
  \State $i^\ast \gets \arg\min_i s_i$
  \State $x^{(g)} \gets c_{i^\ast}$ if $s_{i^\ast} < s(x^{(g-1)})$, else $x^{(g-1)}$ \Comment{keep only the best-so-far}
\EndFor
\State \Return $x^{(G)}$ \Comment{$G\!\times\!N = 1{,}600$ diffusion calls per finished design}
\end{algorithmic}
\end{algorithm}

\subsection{Stage 3: position-value model}
\label{sec:agent}
Stage 3 replaces Stage 2's random mask placement with a learned ranking of candidate positions.
The value of editing position $p$ in image $x$, read at quantile fraction $\tau$, is modelled as a
frozen linear term plus a $\tau$-conditioned residual:
\begin{equation}
  Q(x,p,\tau) \;=\; \underbrace{\beta^{\top}\phi_5(x,p)}_{\text{Ridge, frozen}} \;+\;
  \underbrace{h_\theta\!\Big(\psi_\theta\big(\phi_5(x,p),\,w\big)\;\odot\;\varphi_\theta(\tau)\Big)}
  _{\text{distributional residual, learned}},
  \label{eq:qmodel}
\end{equation}
where $\odot$ is the elementwise product,
\begin{equation}
  \varphi_\theta(\tau) \;=\; \mathrm{ReLU}\!\Big(W_\varphi\big[\cos(\pi k\tau)\big]_{k=1}^{64}
  + b_\varphi\Big),
  \qquad \tau\sim\mathcal{U}(0,1)\ \text{during training}.
  \label{eq:tauembed}
\end{equation}
Here $\phi_5$ is the five-dimensional geometric feature vector described below and $\beta$ is fit
once by ordinary ridge regression ($\alpha{=}100$, $N{=}31{,}581$ position samples on mask $m24$)
and then frozen, i.e.\ excluded from all gradient updates. Here $w$ denotes the policy's
conditioning input, distinct from the fixed scoring-objective weight $w_{\text{corner}}$ of the
Score equation above; the gap between the two is the subject of the supplementary material
(\S\ref{app:cornerweight}). The
learned part is three small
networks: $\psi_\theta$ embeds the geometric features together with the corner weight $w$,
$\varphi_\theta$ is IQN's quantile embedding --- a fixed cosine basis of 64 terms followed by a
learned linear map and a ReLU, so that $\tau$ enters the network rather than merely indexing its
output --- and $h_\theta$ maps their elementwise product back to the scalar improvement value.
Only $\theta$ is trained, by regressing Eq.~\eqref{eq:qmodel} on empirical position-outcome
distributions ($n_{\text{cand}}{=}8$ pre-generated candidate edits per position, \S\ref{sec:trainconfig}) under the
quantile Huber (pinball) loss
\begin{align}
  \mathcal{L}(\theta) &\;=\; \mathbb{E}_{\tau\sim\mathcal{U}(0,1)}
  \left[\frac{1}{n_{\text{cand}}}\sum_{i=1}^{n_{\text{cand}}} \rho_\tau^\kappa\big(y_i(x,p) - Q(x,p,\tau))\right], \label{eq:pinball}\\
  \rho_\tau^\kappa(u) &\;=\; \big|\tau - \mathbb{1}\{u<0\}\big|\cdot
  \begin{cases} \tfrac12 u^2 & |u|\le\kappa \\ \kappa(|u|-\tfrac12\kappa) & \text{otherwise} \end{cases} \label{eq:rho}
\end{align}
where $n_{\text{cand}}{=}8$ is the number of offline candidate edits pre-generated per position,
$y_i(x,p)$ is the realised improvement of its $i$-th draw at $(x,p)$, $\kappa{=}1$, and the outer
expectation is approximated with 32 samples of $\tau$ per
minibatch: supervised regression on a fixed offline label set, with no bootstrapped target and no
environment interaction during training.

\paragraph{The five geometric features $\phi_5$} Each candidate mask window is summarised by the
five scalars of Table~\ref{tab:features}, all computed from the skeleton alone and therefore
available without invoking the diffusion model or the surrogate.

\begin{table}[h]
\caption{The five geometric path features $\phi_5$ and the frozen ridge fit. Pearson $r$ is the
univariate correlation of each feature with the realised improvement; $\beta$ is its coefficient
in the joint ridge fit ($\alpha{=}100$, $N{=}31{,}581$ position samples, mask $m24$, intercept
$b{=}0.01274$). The boundary-crossing count is written $n_b$, matching the F5 panel of
Fig.~\ref{fig:pipeline}, to avoid collision with the ridge coefficient vector $\beta$ of
Eq.~\eqref{eq:qmodel}.}
\label{tab:features}
\begin{center}
\begin{tabular}{cllrr}
\toprule
\# & Symbol & Feature & Pearson $r$ & $\beta$ \\
\midrule
F1 & $d_c$   & Distance from mask centre to image centre    & $+0.0388$ & $+0.00475$ \\
F2 & $d_\gamma$ & Distance from mask centre to nearest path & $+0.0237$ & $+0.00360$ \\
F3 & $n_e$   & Number of broken-path endpoints in the window & $+0.0241$ & $+0.00168$ \\
F4 & $l_p$   & Length of existing path inside the window     & $+0.0089$ & $+0.00139$ \\
F5 & $n_b$   & Number of path crossings of the mask boundary & $-0.0107$ & $-0.00063$ \\
\bottomrule
\end{tabular}
\end{center}
\end{table}

Four of the five promote improvement; the boundary-crossing count $n_b$ suppresses it, since a
location crossed by many paths risks severing an existing connection when edited. Two properties
of this fit matter below. First, the signs are individually interpretable, so the frozen term
works as a usable prior instead of an opaque initialisation. Second, the univariate correlations
are all below $0.04$ in magnitude, meaning the linear term explains almost none of the
per-position variance --- yet ranking positions by it alone already wins against random editing
in five of the seven mask configurations. That count overstates the evidence, for the reasons set
out in the supplementary material (\S\ref{app:arch}): the seven configurations yield only five
distinct ridge-only rollouts, and no individual cell is significant. Section~\ref{sec:discussion}
returns to this gap
between per-position accuracy and deployed ranking quality.

\paragraph{Architectures compared} \textbf{IQN} residual as above. \textbf{FQF}
\citep{fqf2019}: the residual additionally learns $\tau$ itself via a fraction-proposal network,
instead of only querying it. \textbf{DQN}: the residual is a scalar $f(\phi_5,w)$ with no $\tau$
input and hence no risk axis --- architecture-matched (same ridge term, same feature encoder
depth) so that any difference from IQN isolates the value of the distributional/risk-conditioned
residual instead of raw capacity. \textbf{Ridge-only}: the frozen linear term alone, no residual
at all, isolating how much of the position-value signal is already linear in $\phi_5$.

\subsection{Training strategy: supervised regression versus sequential RL}
\label{sec:whynotrl}
An earlier version of this system trained a distributional critic end-to-end by
temporal-difference (TD) learning; that formulation was abandoned in favour of the design below.
Because an offline pool of labelled edit outcomes is available and each edit can be evaluated
independently, we keep only IQN's \emph{query interface} --- conditioning a value network on
$\tau$ --- and fit every component by supervised regression on a fixed offline label set, with no
bootstrapped target, $n$-step return, or replay buffer. \S\ref{sec:tdvsup} verifies this choice
against a controlled TD comparison on the same architecture and data.

\subsection{Deployment policy}
\label{sec:policy}
Deployment is a single deterministic forward pass, not a sequential decision process. For one
held-out image, \emph{every} candidate position $p\in\{1,\dots,P\}$ for the active mask size is
scored at once (full enumeration, never a sample), and each position is reduced to one scalar by
averaging Eq.~\eqref{eq:qmodel} over a risk band:
\begin{equation}
  \bar Q_{\text{risk}}(p) \;=\; \frac{1}{|\mathcal{T}_{\text{risk}}|}
  \int_{\mathcal{T}_{\text{risk}}} Q(x,p,\tau)\,\mathrm{d}\tau ,
  \qquad
  \pi \;=\; \operatorname*{argsort\,desc}_{p=1,\dots,P}\ \bar Q_{\text{risk}}(p).
  \label{eq:policy}
\end{equation}
The band is one of three fixed $\tau$-windows,
$\mathcal{T}_{\text{averse}}{=}[0.001,0.25]$, $\mathcal{T}_{\text{neutral}}{=}[0.01,0.99]$,
$\mathcal{T}_{\text{seeking}}{=}[0.75,0.999]$; the integral is evaluated as a uniform average over
32 quantile points per window. The permutation $\pi$ that results is the priority list, and the
policy visits positions in that fixed order for 20 steps; revisits are structurally impossible
and no information flows from step to step. We call this the \textbf{static greedy ranking
policy}. Only the band, not the shape of the predicted distribution, enters the
decision: two positions with very different predicted distributions but equal band averages are
interchangeable to $\pi$. This is the sense in which the policy uses the distributional output as
a ranking device (\S\ref{sec:calibration}). Both the risk band $\mathcal{T}_{\text{risk}}$ and the corner weight $w$ are inference-time
inputs: changing either and re-scoring produces a new priority list with \emph{no
retraining} (\S\ref{sec:mechanism}, supplementary \S\ref{app:cornerweight}).

Algorithms~\ref{alg:stage2} and \ref{alg:stage3} make the budget contrast between the two stages
explicit: Stage 2 repeatedly scores $N$ freshly generated candidates per generation and keeps only
the best-so-far design, while Stage 3 scores every candidate position once, in a single deterministic
pass, and never re-scores a visited position.

\begin{algorithm}[H]
\caption{Stage 3 --- static greedy ranking policy}
\label{alg:stage3}
\begin{algorithmic}[1]
\Require held-out bare skeleton $x$ ($41{\times}33$); trained $\{\beta,\theta\}$; risk band $\mathcal{T}_{\text{risk}}$; corner weight $w$; 20 steps
\For{$p = 1, \dots, P$}
  \State $\bar Q_{\text{risk}}(p) \gets$ average of Eq.~\eqref{eq:qmodel} over $\mathcal{T}_{\text{risk}}$ (32 quantile pts), from $\phi_5(x,p)$ \Comment{no diffusion call}
\EndFor
\State $\pi \gets \operatorname*{argsort\,desc}_{p} \bar Q_{\text{risk}}(p)$ \Comment{static priority list, computed once}
\For{$t = 1, \dots, 20$}
  \State $p \gets \pi[t]$
  \State $x' \gets$ DDPMInpaint$(x, p)$ \Comment{same operator as Alg.~\ref{alg:stage2}: $41{\times}33{\to}41{\times}41{\to}64{\times}64$, 10-step, $\le\!5$ attempts}
  \State $x \gets x'$ if $x'$ passes single-stroke $+$ inlet/outlet validity (\S\ref{sec:score}), else $x$ unchanged
\EndFor
\State \Return $x$ \Comment{scored via routing $\to 48{\times}35$; $20\!\times\!1 = 20$ diffusion calls, no re-scoring}
\end{algorithmic}
\end{algorithm}

\section{Experimental Setup}
\label{sec:setup}

\subsection{Data generation (ground truth)}
\label{sec:data}
Ground-truth labels come from CFD: OpenFOAM \texttt{chtMultiRegionFoam} (transient conjugate heat
transfer, laminar), two regions (coolant fluid + aluminum-grade solid), the same open-source
solver \citet{kimhan2023} couple to their DQN cooling-channel agent (validated there against a
commercial Ansys cross-check), which motivates our own analogous validation practice
(\S\ref{app:cfd} of the supplementary material). Layouts are generated by a mirror-symmetric maze and a longest-path
depth-first search, meshed with \texttt{blockMesh} + \texttt{topoSet} using our own OpenFOAM
case-generation pipeline, with $T_{\max}$, $T_{\min}$,
$T_{\text{avg}}$, $T_{\text{uni}}$ and $\Delta p$ extracted by function objects (the resulting
labelled data's availability is discussed in \S\ref{sec:repro}). Each accepted skeleton is aligned to predefined inlet/outlet
positions by one of two routing conventions (Type A/B); both are used to train the surrogate, but
Type B is fixed for all editing and evaluation. A layout is accepted only if its extracted path
exceeds a length threshold, and that threshold separates the two populations used here.
The \textbf{training set} pools the layouts accepted at thresholds above 300 (2{,}000 layouts,
$4{,}000$ image--label pairs across both routing types). The \textbf{held-out set} is generated
at the stricter threshold of 360 (100 layouts) and supports every judgement in
\S\ref{sec:results}. Because a higher threshold admits only longer channels, the held-out
population is not a random split of the training population but a deliberately harder sample
drawn from the same generator: its mean $T_{\text{avg}}$ and $\Delta p$ both sit roughly $2.8$ standard
deviations above the training mean, and only $34$--$37\%$ of held-out layouts fall inside the
training population's central $95\%$ interval on those two quantities. Held-out numbers in
\S\ref{sec:results} are therefore measured on the upper tail of the distribution the surrogate
was fitted to, not at its centre.
\begin{figure}[htbp]\centering
\includegraphics[width=0.92\textwidth]{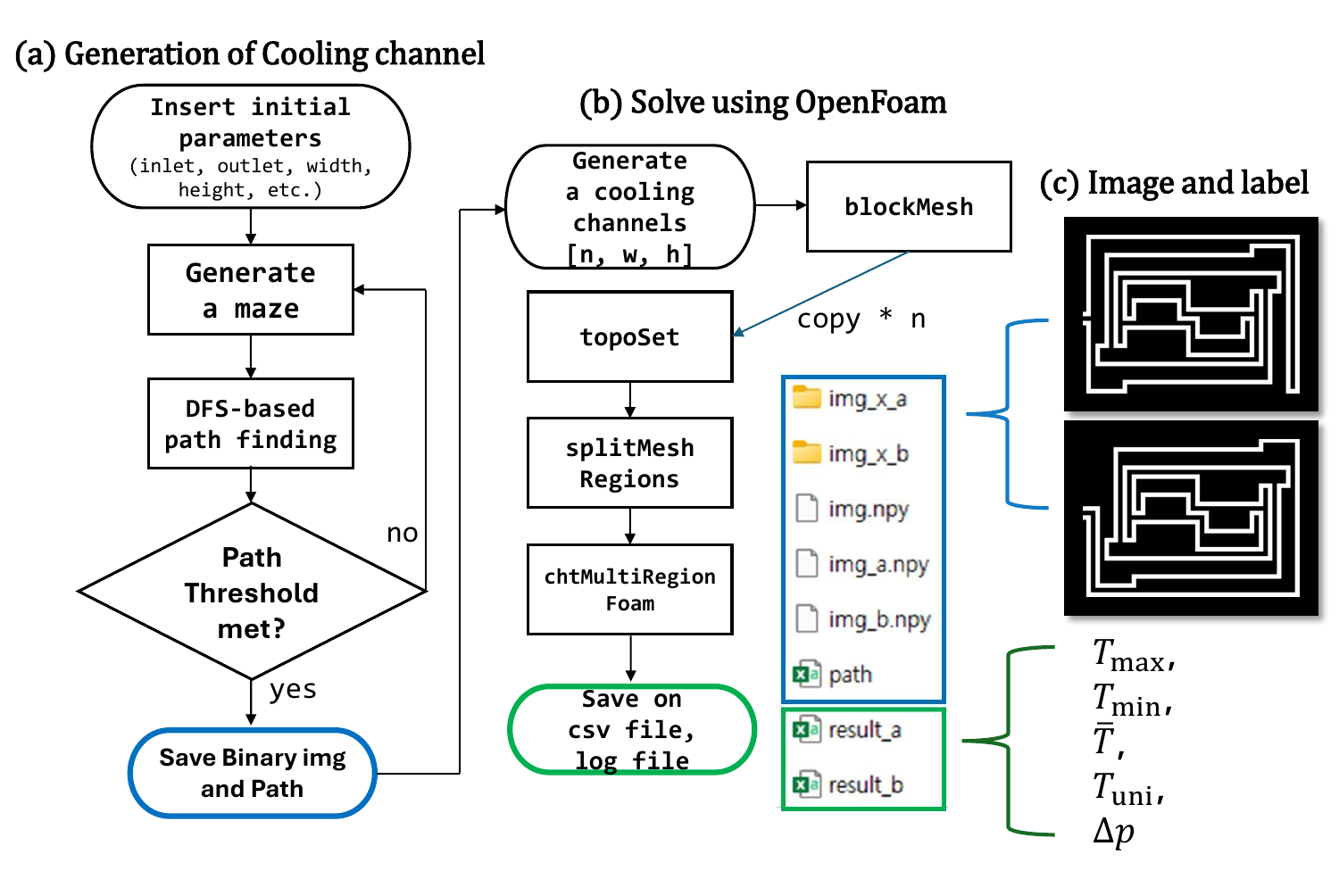}
\caption{Dataset preparation. (a) Layout generation: from the initial parameters (inlet, outlet,
channel width and height) a maze is generated and a depth-first search extracts the longest path;
a layout whose path fails the length threshold is discarded and regenerated, and only accepted
layouts are saved as a binary skeleton with its path. (b) CFD solution: each accepted layout is
replicated into $n$ cases and solved through \texttt{blockMesh} $\rightarrow$ \texttt{topoSet}
$\rightarrow$ \texttt{splitMeshRegions} $\rightarrow$ \texttt{chtMultiRegionFoam}, with the
results written to CSV. (c) Image and label: each layout is stored both as the bare skeleton and
as its two routing assemblies (\texttt{img\_a}, \texttt{img\_b}), paired with the five response
quantities $T_{\max}$, $T_{\min}$, $T_{\text{avg}}$, $T_{\text{uni}}$ and $\Delta p$ extracted per
assembly.}
\label{fig:dataprep}
\end{figure}

\subsection{Surrogate model}
\label{sec:surrogatedetail}
The network is an EfficientNetV2-S backbone \citep{tan2021effnetv2} with a five-output regression
head, trained on the
\emph{assembled} layout (inlet/outlet routing included), predicting
$[T_{\max}, T_{\min}, T_{\text{avg}}, T_{\text{uni}}, \Delta p]$; its accuracy on the labelled
population appears in \S\ref{sec:stage1}. The assembled-vs.-bare-skeleton distinction
matters: an earlier iteration of the pipeline fed the
surrogate a geometrically incomplete (bare-skeleton) input at scoring time, which produced
numerically plausible but physics-uncorrelated scores ($r{\approx}0.04$) without raising any
error and silently inverted which arm appeared to win. The lesson has stayed with us:
any pipeline that couples a learned evaluator to a generative editor should treat
input-convention agreement with the training data as a release gate, not an assumption.

\subsection{Training configuration}
\label{sec:trainconfig}
For each candidate edit position, 8 candidate completions are pre-generated offline (best-of-8
per generation in the diffusion search stage) and serve as the empirical outcome distribution on
which the residual is regressed. The residual network is compact, roughly a 3-layer MLP, so a
full 300-epoch training run takes about 105--195 seconds; Table~\ref{tab:hyperparams} gives its
architecture and optimization hyperparameters in full.
All results below use
three training seeds; per-mask effects are pooled across these three
seeds by inverse-variance fixed-effect meta-analysis, and a further random-effects meta-analysis
pools across the seven mask configurations. No additional training seeds were run beyond these
three, a pre-registered decision rather than a stopping-early choice.

\begin{table}[h]
\caption{Residual network ($\psi_\theta,\varphi_\theta,h_\theta$ of Eq.~\eqref{eq:qmodel}) architecture and optimization hyperparameters. The frozen ridge term $\beta^{\!\top}\phi_5$ is fit separately (\S\ref{sec:agent}) and is not part of this table.}
\label{tab:hyperparams}
\begin{center}
\begin{tabular}{ll}
\toprule
Field & Value \\
\midrule
Feature encoder $\psi_\theta$ & two 128-unit layers (ReLU), input $[\phi_5(x,p), w]$ \\
Quantile embedding $\varphi_\theta$ & 64-term cosine basis $\to$ learned linear map $\to$ ReLU (Eq.~\eqref{eq:tauembed}) \\
Combination & elementwise product $\psi_\theta \odot \varphi_\theta(\tau)$ \\
Output head $h_\theta$ & linear readout to the scalar improvement value \\
Loss & quantile Huber / pinball, $\kappa{=}1$ (Eqs.~\eqref{eq:pinball}--\eqref{eq:rho}) \\
$\tau$ sampling & $\tau\sim\mathcal{U}(0,1)$, 32 samples per minibatch \\
Candidates per position & $n_{\text{cand}}{=}8$ (offline pre-generated, best-of-8) \\
Optimizer & AdamW, learning rate $10^{-3}$, weight decay $10^{-4}$ \\
Batch size & 16 \\
Epochs & 300, no learning-rate schedule \\
Training seeds & 3 (fixed-effect pooled; \S\ref{sec:c1}) \\
Wall-clock (full run) & 105--195\,s \\
\bottomrule
\end{tabular}
\end{center}
\end{table}

\subsection{Evaluation protocol}
\label{sec:protocol}
Evaluation is \emph{paired}: the trained policy and a random-position baseline are scored on the
\emph{same} 100 held-out images (a fixed held-out index seed, seed \#1), each rolled out for 20 steps with
$n_{\text{cand}}{=}1$ (one new sample scored per step, no best-of-$N$ at deployment). Statistics
are per-image paired $t$-tests (image $=$ independent unit, no episode pooling), with all
81--371 candidate positions of the active mask scored, sorted, and visited deterministically per
\S\ref{sec:policy}. One property of the task shapes this protocol: a candidate edit is accepted
only if the resulting layout still satisfies the single-stroke connectivity constraint of
\S\ref{sec:score}, one continuous channel so that the coolant path remains a single
manufacturable route, and a masked region cannot always be reconnected that way. When no
sampled candidate at the visited position satisfies the constraint, the step leaves the layout
unchanged. Under the project's pre-registered statistical policy, headline numbers
are reported at raw (uncorrected) $p$-values \emph{when} the compared cell was fixed before the
data were seen (one representative condition, IQN, risk-seeking, $w{=}0.001$, per mask); any
number selected \emph{after} seeing results (e.g.\ ``the best cell for this mask'') is flagged as
post-hoc and never supports a headline claim, because a companion analysis
found that post-hoc per-mask winners do not reproduce across training seeds and can flip sign
entirely.

\section{Results}
\label{sec:results}
Results follow the three stages of \S\ref{sec:system}. \S\ref{sec:stage1} establishes that the
surrogate reproduces CFD on the labelled population, the licence for using it as the
objective at all. \S\ref{sec:stage2} reports what iterative inpainting search achieves with
\emph{no} learned policy, the capability the learned policy has to improve on.
\S\ref{sec:stage3} covers Stage 3, the learned editing policy, the contribution of
this paper. Stage 2 and Stage 3 answer different questions and are never pooled: Stage 2 spends
many generations of best-of-$N$ search on one design, whereas Stage 3 makes one deterministic
pass with a single candidate per step (\S\ref{sec:protocol}).

\subsection{Stage 1: labelled population and surrogate fidelity}
\label{sec:stage1}
\begin{figure}[htbp]\centering
\includegraphics[page=1,width=\textwidth]{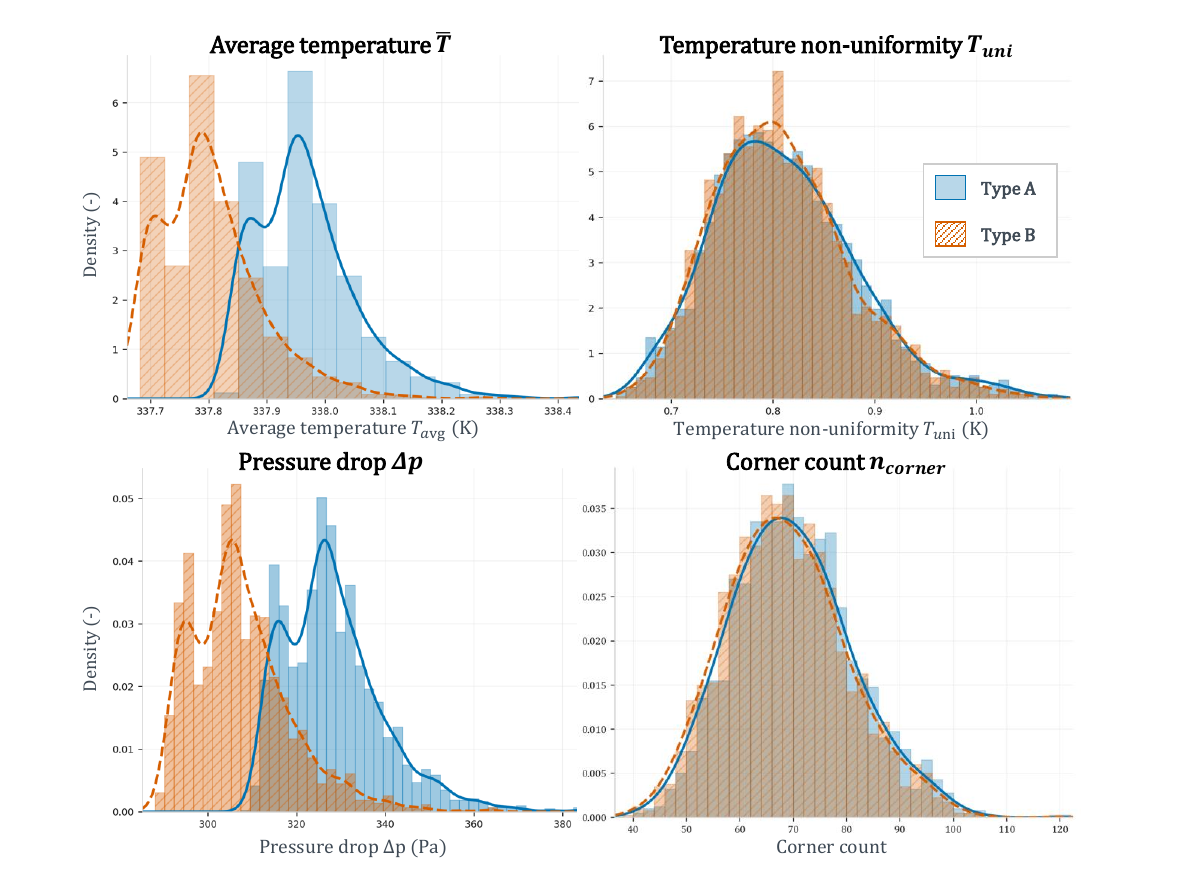}
\caption{Stage 1. Distribution of the four CFD-labelled response quantities over the generated
population ($N{=}2{,}000$ layouts, each assembled under both routing conventions). The two
routing types differ in \emph{level} on the flow-coupled quantities ($T_{\text{avg}}$, $\Delta p$)
but almost coincide on $T_{\text{uni}}$ and corner count, which depend on the channel skeleton
rather than on how it is connected to the inlet and outlet.}
\label{fig:datadist}
\end{figure}
\begin{figure}[htbp]\centering
\includegraphics[page=2,width=\textwidth]{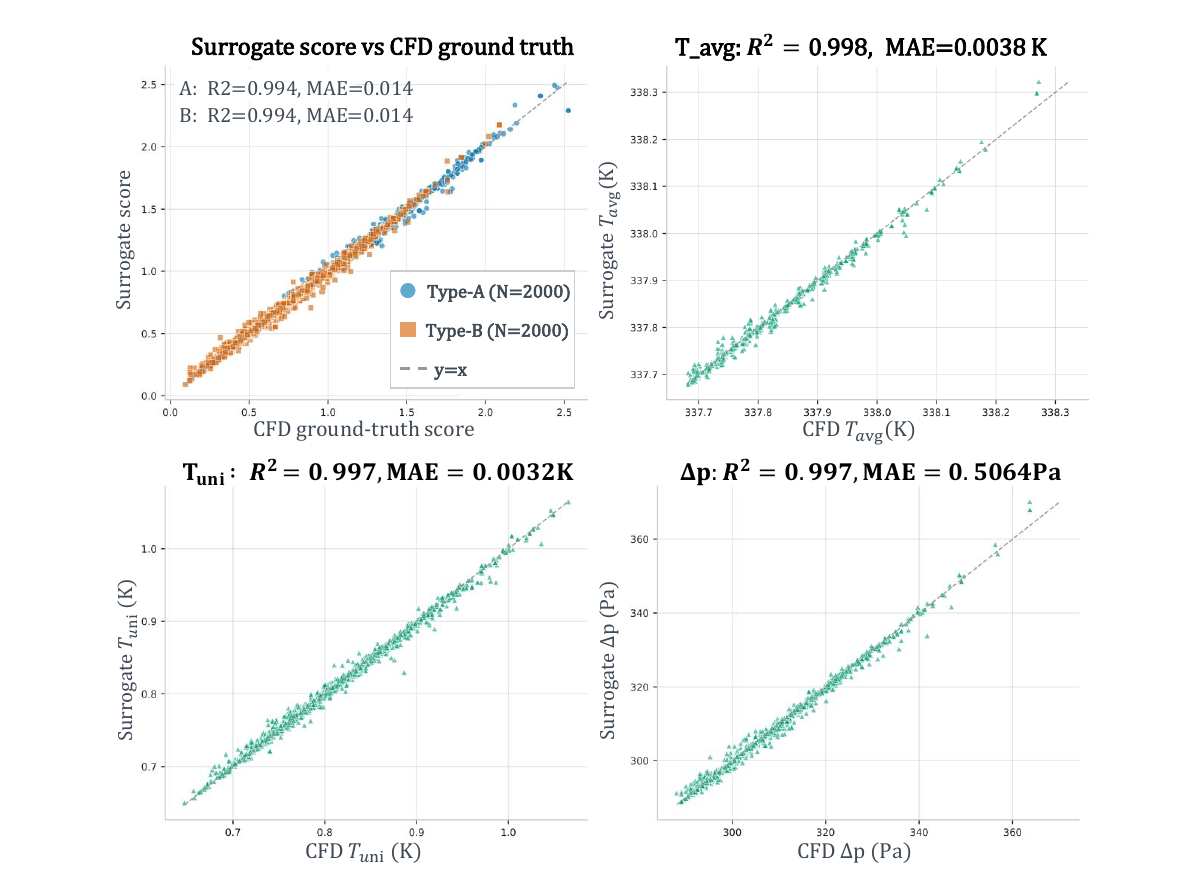}
\caption{Stage 1. Surrogate $\leftrightarrow$ CFD agreement on the labelled population. Top left:
the composite objective, for both routing types. Remaining panels: the individual physical
responses. The dashed line is $y{=}x$.}
\label{fig:cfd}
\end{figure}
Fig.~\ref{fig:datadist} shows the labelled population the surrogate is trained on. The
optimization problem it defines is a narrow one: the entire population spans about $0.6\,$K in
$T_{\text{avg}}$, so the objective is governed by relative rather than absolute differences, and
a scoring model has to resolve the population at that scale to be useful. The two routing
conventions shift $T_{\text{avg}}$ and $\Delta p$ noticeably while leaving $T_{\text{uni}}$ and
corner count nearly unchanged; this is why the routing convention is held fixed (type-B) for all
editing and evaluation (\S\ref{sec:data}) instead of being treated as a free variable.

The surrogate resolves the population at that scale. It reproduces the composite CFD objective
with $R^2{=}0.994$ and MAE $0.014$ for both routing types (Fig.~\ref{fig:cfd}, top left;
$r{=}0.997$), and the individual responses are recovered at least as well: $T_{\text{avg}}$
$R^2{=}0.998$ with MAE $0.0038\,$K, $T_{\text{uni}}$ $R^2{=}0.997$ with MAE $0.0032$, and
$\Delta p$ $R^2{=}0.997$ with MAE $0.51\,$Pa. The $T_{\text{avg}}$ error sits roughly two orders of
magnitude below the population spread, which is what allows the surrogate to stand in for
CFD inside the Stage 2 and Stage 3 loops. Whether that fidelity carries over to \emph{edited}
designs is a different and stronger claim, tested separately in \S\ref{sec:cfd_verify}.

\subsection{Stage 2: best-of-$N$ optimization by inpainting, without a learned policy}
\label{sec:stage2}
\begin{figure}[htbp]\centering
\includegraphics[page=1,width=\textwidth]{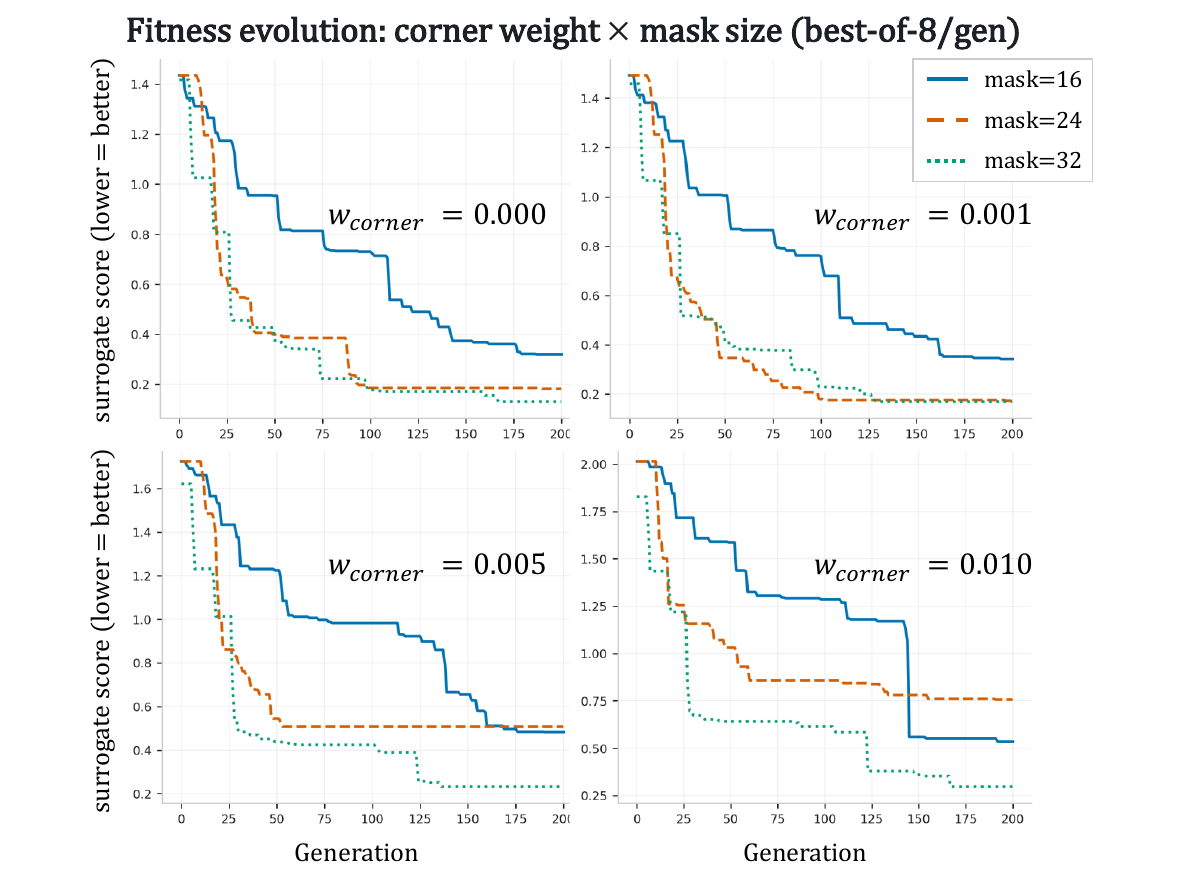}
\caption{Stage 2. Fitness evolution under best-of-8 inpainting search, for each corner weight
$w_{\text{corner}}$ (panels) and mask size (curves). Lower is better. Search is monotone by
construction (the best-so-far design is retained), so these curves show how quickly, not whether,
the objective falls.}
\label{fig:stage2curve}
\end{figure}

\begin{figure}[htbp]\centering
\includegraphics[page=1,width=\textwidth]{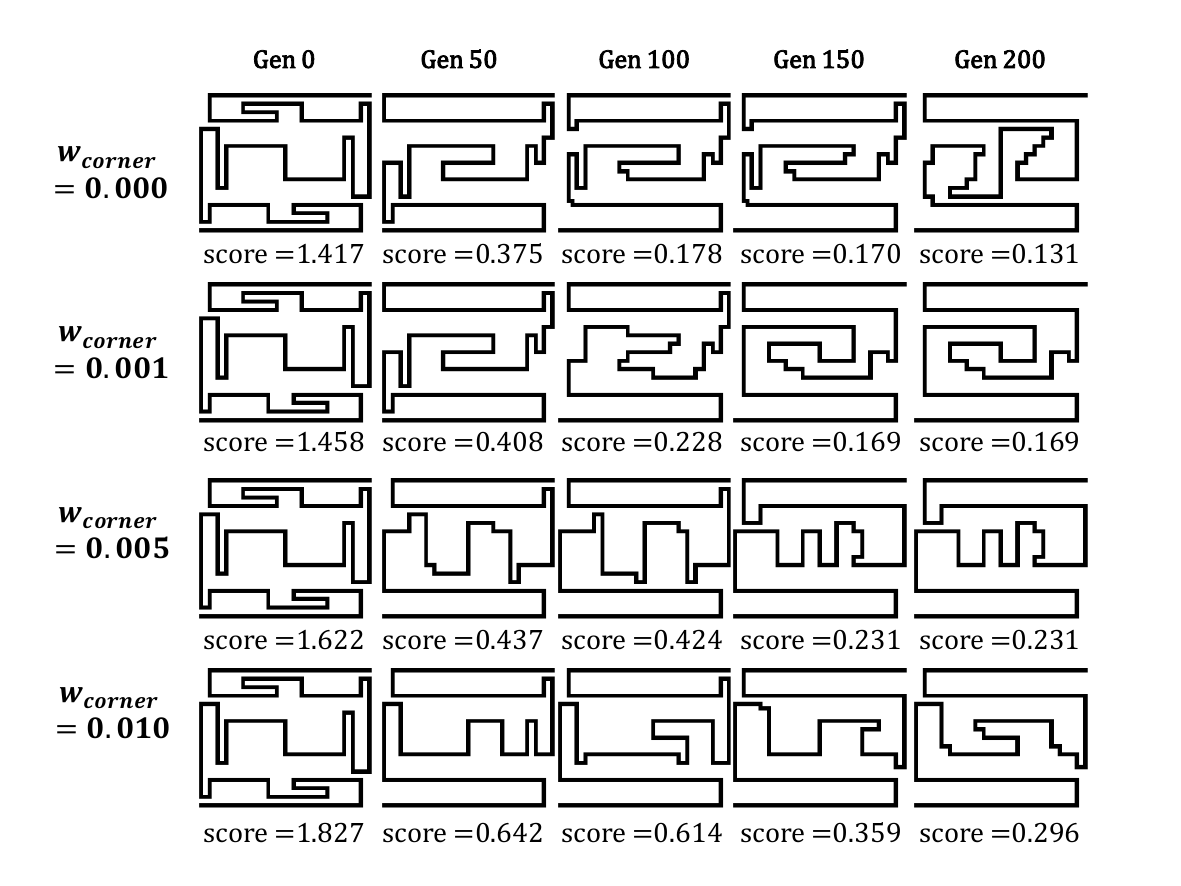}
\caption{Stage 2. Channel layouts at generations 0, 50, 100, 150 and 200 for each corner weight
($\text{mask}{=}32$), with surrogate scores. Raising $w_{\text{corner}}$ visibly suppresses
staircase-like corner structure, which is what the term is intended to penalize, at the cost of a
higher achieved score under the composite objective.}
\label{fig:stage2grid}
\end{figure}

This stage establishes the premise the learned policy depends on: that editing an existing
layout with the inpainting operator improves it at all, and by how much when the number of
edits is unconstrained. Stage 3 then takes up a different question, where to edit first when
the budget is small, at a budget two orders of magnitude lower; the two stages answer different
questions and are not competing methods. Here we measure what the operator achieves
under an unguided best-of-$N$ search: at each generation, $N{=}8$ candidate edits are proposed at randomly
placed masks, all are scored by the surrogate, and the best is retained. Fig.~\ref{fig:stage2curve}
shows the resulting fitness evolution over 200 generations for every combination of mask size and
corner weight, and Fig.~\ref{fig:stage2grid} shows the corresponding layouts.
Two observations matter for the rest of the paper. First, unguided search is already strong: over
200 generations the surrogate score falls from $1.417$ to $0.131$ at $w_{\text{corner}}{=}0$, and
the corresponding physical quantities improve consistently ($T_{\text{avg}}$ $337.998\rightarrow
337.656\,$K, $T_{\text{uni}}$ $0.9165\rightarrow0.6566$, $\Delta p$ $334.01\rightarrow288.96\,$Pa,
corner count $49\rightarrow35$, for $\text{mask}{=}32$). A learned policy therefore has to justify
itself against a baseline far from trivial --- though, as set out above, it is not asked to
beat this number. Stage 2 spends $200$ generations
$\times\,8$ candidates $=1{,}600$ diffusion calls to produce one finished design. Stage 3 spends
$20$ steps $\times\,1$ candidate $=20$ calls, an $80\times$ smaller budget, and its ranking is
computed by a single forward pass that invokes no diffusion at all. The question Stage 3 answers
is therefore not ``can a learned policy beat exhaustive search'' but ``when the edit budget is
small, does knowing \emph{where} to edit beat editing at random.'' Second, mask size
dominates convergence speed:
$\text{mask}{=}16$ is consistently the slowest to converge across all four corner weights,
foreshadowing the mask-size effect that reappears in Stage 3 (Table~\ref{tab:mechanism}).

An independent CFD check of Stage 2 designs confirms the direction of improvement for the two
larger masks (paired, $N{=}30$ designs per mask: $m24$ $p{=}0.009$, $m32$ $p{=}0.012$), but not
for $m16$ ($p{=}0.240$), where the surrogate-predicted gains do not reach physical significance at
this sample size. Two caveats attach to these curves. They are surrogate scores, and late
generations push designs toward the edge of the range the surrogate was trained on, where its
error grows (\S\ref{sec:cfd_verify}); and because best-of-$N$ retains the best-so-far design, the
curves cannot rise, so they should be read as convergence rates, not as evidence that
every generation improved the design.

\subsection{Stage 3: the learned editing policy}
\label{sec:stage3}
Stage 3 asks whether ranking edit positions with a learned value model beats an unguided
ordering under the same edit budget. The subsections below answer that in four steps: the
pre-registered held-out comparison, and how far its effect size moves across independent rollout
seeds (\S\ref{sec:c1}); which axis the advantage actually lives on ---
the inference-time risk level rather than the architecture --- what it costs in seed-to-seed
consistency, and how it degrades when the conditioning value disagrees with the objective being
scored (\S\ref{sec:mechanism}--\S\ref{sec:seedspread}); whether the surrogate-level advantage
survives CFD (\S\ref{sec:cfd_verify}); and what the learned quantile function is, and is not,
good for (\S\ref{sec:tdvsup}).

\subsubsection{Held-out improvement across seven mask configurations}
\label{sec:c1}
\begin{table}[h]
\caption{Pre-registered headline condition (IQN, risk-seeking, $w{=}0.001$) vs.\ random, per
mask, held-out $N{=}100$, 20-step paired rollout, fixed-effect pooling over three training
seeds, a single rollout seed (seed \#1). ``mm'' denotes the union of all three single masks and
``pair($a$,$b$)'' the union of two. Action space $P$ = number of candidate edit
positions for that mask configuration.}
\label{tab:c1}
\begin{center}
\begin{tabular}{lcccc}
\toprule
Mask & $P$ & $\Delta$ (agent $-$ random) & $z$ & $p$ \\
\midrule
m16                & 169 & $+0.0441$ & $2.59$ & $0.0096$ \\
m24                & 121 & $+0.0545$ & $2.47$ & $0.0136$ \\
m32                & 81  & $+0.0415$ & $1.86$ & $0.063$ \\
mm (16+24+32)      & 371 & $+0.0876$ & $4.03$ & $5.5\times10^{-5}$ \\
pair(16,24)        & 290 & $+0.1389$ & $6.51$ & $7.8\times10^{-11}$ \\
pair(24,32)        & 202 & $+0.0650$ & $3.16$ & $0.0016$ \\
pair(16,32)        & 250 & $+0.0633$ & $2.75$ & $0.0060$ \\
\bottomrule
\end{tabular}
\end{center}
\end{table}
The pre-registered headline condition beats random editing in \textbf{six of seven} mask
configurations (Table~\ref{tab:c1}); the seventh, $m32$, misses at $p{=}0.063$ and is reported as
a borderline null instead of being omitted. Pooling all seven conditions by random-effects
meta-analysis gives a mask-averaged effect of $\Delta{=}+0.0674$ ($95\%$ CI $[0.0402,0.0946]$,
$z{=}4.86$, $p{=}1.2\times10^{-6}$), which agrees with the per-mask table by construction (same
condition definition) and is the number we quote as ``the agent beats random editing on
average.'' Two features of this pooling belong alongside it. The heterogeneity is
substantial ($I^2{=}69.4\%$): the seven mask configurations do not share one common effect size,
and the pooled value averages over configurations that genuinely differ, as
Fig.~\ref{fig:seedspread_pct} shows directly as well. The seven effects are also not seven
independent samples, since every configuration is evaluated on the \emph{same} fixed 100 held-out
designs and their sampling errors are therefore correlated. The mixed-effects model of
\S\ref{sec:mechanism}, which pools all $15{,}200$ rows with an image random intercept and
models that shared structure explicitly, reaches the same conclusion, and we treat it
as the check on this one. That model reports only main effects (Table~\ref{tab:mechanism}); we
do not add a mask$\times$risk interaction term here because the question it answers is whether
architecture identity retains a main effect once mask and risk are controlled for, not how the
risk effect varies mask by mask, which Table~\ref{tab:c1} and Fig.~\ref{fig:frontier_risk} already
report directly. The action-space size $P$ scales with mask footprint as expected
($P_{16{+}24{+}32}=P_{16}+P_{24}+P_{32}$ exactly, confirming the multi-mask action space is the
union of the single-mask spaces).

\label{sec:rolloutseed}
The Table~\ref{tab:c1} headline number, like the rest of Table~\ref{tab:c1}, is measured at a
single, pre-registered DDPM rollout seed (seed \#1). Because the environment transition is
itself stochastic (\S\ref{sec:stage2method}), we additionally measured the headline configuration
(pair(16,24)) at five independent rollout seeds (seeds \#1--\#5). The pre-registered seed, one of the five, gives
$\Delta{=}+0.1317$; across the five the range is $[+0.0537, +0.1317]$
(mean $+0.094$, s.d.\ $0.031$), and the seed-averaged, per-image paired comparison gives
$t{=}3.72$, $p{=}3.3\times10^{-4}$, a 63/100 win rate. The corresponding spread across the three
\emph{training} seeds, a separate axis, is given in the supplementary material
(\S\ref{app:robustness}).
The effect is present at every rollout seed tested; what varies is its size. Table~\ref{tab:c1}
--- the pre-registered protocol, one rollout seed --- remains the reporting
basis throughout, so that every configuration in this paper is quoted on the same footing, with
the seed range reported beside it rather than folded into the headline: on this
configuration the pre-registered seed is the largest of the five, and the seed-averaged estimate
is $+0.094$, some $30\%$ below the single-seed value measured at the same training seed
($+0.1317$). Extending the same measurement to a consecutive block of 24 rollout seeds (seeds \#1--\#24) leaves
that picture unchanged, with a positive effect in all 24 and a 24-seed average of
$+0.083$ ($p{=}1.8\times10^{-4}$), consistent with the five-seed spread reported here. The same single-rollout caveat applies to the other six
mask configurations and to the pooled estimate above.

\subsubsection{Factor decomposition of the headline effect}
\label{sec:mechanism}
\begin{table}[h]
\caption{Mixed-effects decomposition of held-out improvement by factor (image random intercept;
reference levels: mask $=m16$, architecture $=$ random baseline, risk $=$ neutral, $w{=}0.001$).
IQN and FQF are estimated from three training seeds at the matched cell (supplementary
\S\ref{app:arch}); DQN and ridge-only rest on a single training seed each.}
\label{tab:mechanism}
\begin{center}
\begin{tabular}{llcc}
\toprule
Factor & Level & Coefficient & $p$ \\
\midrule
Mask     & (all levels, vs.\ $m16$)   & $+0.079$ to $+0.139$ & all $<10^{-20}$ \\
Risk     & seeking                     & $+0.0471$ & $9.5\times10^{-15}$ \\
Risk     & averse                      & $+0.0118$ & $0.053$ (borderline) \\
$w$      & $0.005$                     & $-0.0202$ & $3.2\times10^{-4}$ \\
$w$      & $0.010$                     & $-0.0319$ & $1.2\times10^{-8}$ \\
Architecture & ridge-only / DQN / IQN / FQF & $+0.001$ to $+0.014$ & all $\ge 0.32$ \\
\bottomrule
\end{tabular}
\end{center}
\end{table}
Table~\ref{tab:mechanism} decomposes the headline effect. Architecture identity, whether the
residual is IQN's, FQF's, DQN's scalar variant or absent entirely (ridge-only), has
\emph{no} significant main effect once mask and risk level are accounted for. The measurable
benefit sits on the risk axis: risk-\emph{seeking} is a large, highly significant effect, risk-
\emph{averse} a borderline positive one, consistent with the per-mask pattern in
\S\ref{sec:c1}, where the winning configuration is risk-seeking in every mask that has a
significant winner. The supplementary material (\S\ref{app:arch}) gives the per-arm comparison
behind the architecture row, together with the full post-hoc grid from which the same
conclusion can be read directly.
One qualification belongs with this row: a targeted three-seed comparison of IQN against FQF at
the matched pre-registered cell, a more powerful test than the pooled factor above because it
pairs the two architectures on the same designs, does separate them, with IQN ahead in five of
seven mask configurations and tied in the other two (supplementary material, \S\ref{app:arch}). The factor model says
architecture is not where the \emph{large} effect lives; it does not say the architectures are
identical, and at the operating point we deploy they are not.
\S\ref{sec:seedspread} measures what is traded away when that dial is turned toward seeking.

\subsubsection{Risk vs.\ seed-to-seed consistency}
\label{sec:seedspread}
\begin{table}[h]
\caption{Within-mask correlations of the seed-to-seed spread with the three factors that can be
set at or before deployment. The seed-spread ratio is random\_std / agent\_std over $K{=}5$ rollout
noise seeds ($N{=}20$ images per condition); $>1$ means the agent's outcomes vary less than
random's. The first and third rows pool the seven mask configurations at all three conditioning
values ($n{=}63$ cells); because the random arm is a single shared baseline within a mask, both
are equivalent to statements about the agent's own spread. The second row is measured against
that spread directly and over the three single-mask configurations ($n{=}27$), since the action
space is a property of the mask configuration and the paired and multi-mask settings do not lie
on the same axis.}
\label{tab:seedspread}
\begin{center}
\begin{tabular}{lcc}
\toprule
Comparison & Spearman $\rho$ & $p$ \\
\midrule
Risk level (averse$\to$neutral$\to$seeking) vs.\ seed-spread ratio & $-0.564$ & $1.5\times10^{-6}$ \\
Mask size vs.\ agent seed spread (single masks) & $+0.833$ & $7.1\times10^{-8}$ \\
Corner weight $w$ (conditioning input) vs.\ seed-spread ratio & $+0.023$ & $0.86$ (n.s.) \\
\bottomrule
\end{tabular}
\end{center}
\end{table}
Moving the risk dial from averse toward seeking increases seed-to-seed variance in every mask
configuration tested (7/7): the agent buys mean improvement at the cost of consistency. Mask size
moves both axes together and in the same direction: across the single-mask configurations a larger
mask brings both a larger mean improvement ($\rho{=}+0.646$, $p{=}2.7\times10^{-4}$) and a larger seed
spread ($\rho{=}+0.833$), measured on the same repeated-seed campaign. Mask size thus sets
a floor on achievable consistency, and the risk level moves within it. The corner weight $w$, by contrast, is \emph{not} correlated with seed spread
at all ($\rho{=}0.023$, n.s.): it moves the mean (Table~\ref{tab:mechanism}) but not the variance
axis. One caveat concerns the level of these ratios. They are measured with the scoring objective held
at $w{=}0.001$ while the conditioning input is swept (supplementary \S\ref{app:cornerweight}); a follow-up
campaign that instead matches the scoring objective to the conditioning value, for the agent arm
only since the random baseline is not rescored, reproduces the direction throughout, but returns
a \emph{lower} ratio in all 42 cells of the grid. Because that follow-up
rescores one of the two arms and not the other, it bears on the level of the ratio rather than on its
ordering. The ordering that Table~\ref{tab:seedspread} rests on is therefore robust, while the
absolute ratio levels reported here should be read as upper estimates.
A second caveat applies to those levels for a different reason: the two arms carry different
sources of seed-to-seed variation. The agent visits a fixed sequence of positions, so its
spread reflects the generator alone, whereas the random arm re-draws its positions at every
seed and its spread therefore also contains the variability of that choice. That is the
comparison a practitioner actually faces, since random editing does land elsewhere each time,
but the level should not be read as a measure of robustness to generator noise. The
correlations in Table~\ref{tab:seedspread} are unaffected by this: within a mask the random arm is a single
shared baseline, so the risk and corner-weight rows are equivalent to statements about the
agent's own spread.

\begin{figure}[!htbp]\centering
\includegraphics[width=\textwidth]{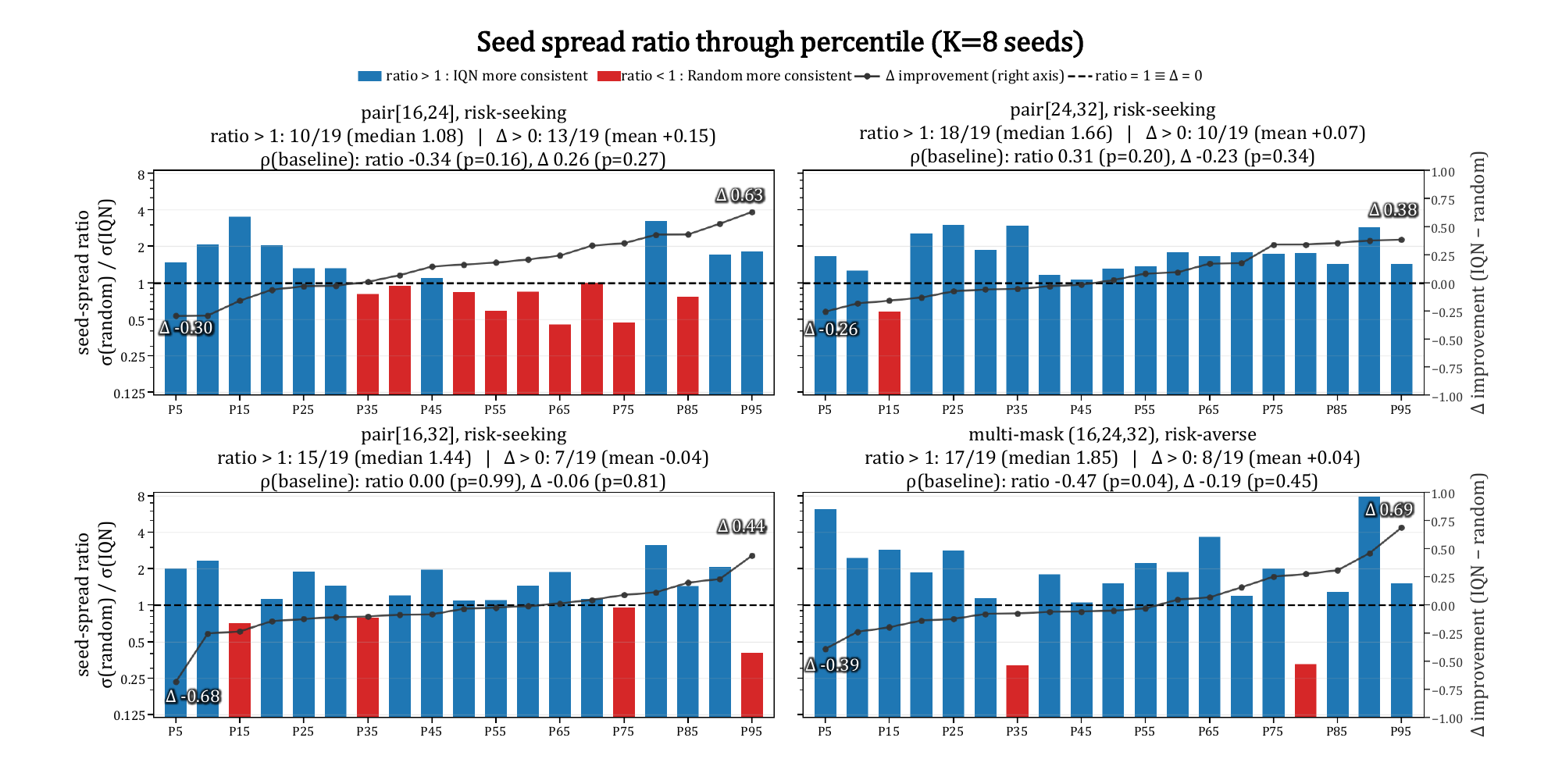}
\caption{Seed-to-seed spread across the held-out improvement range. Held-out images are ordered by
improvement and divided into 19 percentile slices. Bars show the seed-spread ratio
$\sigma_{\mathrm{random}}/\sigma_{\mathrm{IQN}}$ across eight independent DDPM rollout seeds, and
the line shows the corresponding improvement $\Delta$. The dashed line indicates ratio $=1$ and
$\Delta=0$.}
\label{fig:seedspread_pct}
\end{figure}

Fig.~\ref{fig:seedspread_pct} restricts to the four mask configurations with a significant mean
effect, panelled by mask composition, pairs first; both the bars and the line for a given panel
come from the same $K{=}8$ rollouts. The two axes are scaled so that ratio${=}1$ and $\Delta{=}0$
coincide on the dashed reference, above which the agent is the better choice under either
encoding --- blue marks a slice where the agent is more consistent, red one where random editing
is, and the grey numbers in each panel's corners give $\Delta$ at the first and last slice. Two
axes that are easily conflated come apart here: pair(24,32) is the most consistent configuration
(ratio${}>{}1$ in 18/19 slices) but wins only 10/19 slices; pair(16,24), the configuration with the
largest mean effect in Table~\ref{tab:c1}, wins the most slices (13/19) while being the weakest on
the consistency axis (10/19); and pair(16,32) is consistent but does not improve (15/19 vs.\
7/19). Across the four configurations the two measures are not correlated (Spearman
$\rho{=}-0.18$, $p{=}0.13$, $n{=}75$ slices). Every configuration contains both favourable and
unfavourable slices, so no configuration performs uniformly well, and the aggregate effects of
Table~\ref{tab:c1} are averages over exactly this kind of mixture. One multi-mask slice
substitutes the nearest edited design in the same ranking, because at that percentile the agent
completes no net edit in any rollout, leaving $\sigma_{\text{IQN}}{=}0$ and the ratio undefined
--- a ratio obtained by not acting is not consistency in any useful sense. The percentile slices
are descriptive only: each slice is a single image and individual points should not be read as
significance tests; all significance claims instead use the pooled $N{=}100$ results of
Table~\ref{tab:c1}.

Fig.~\ref{fig:seedspread_pct} shows that this consistency behaviour is not a property of a few
favourable designs. It holds across the full improvement range of the held-out set and in every
mask configuration with a significant mean effect, and it makes visible where it does
\emph{not} hold, so the single-design illustrations that follow can be read against the
population they are drawn from rather than in place of it. Because the same checkpoint is also
conditioned on $w$ at inference (\S\ref{sec:policy}), we additionally swept the conditioning value
away from the scoring objective it was trained under; the mismatch degrades performance, and the
detailed sweep is reported in the supplementary material (\S\ref{app:cornerweight}).

\begin{figure}[!htbp]\centering
\includegraphics[width=\textwidth]{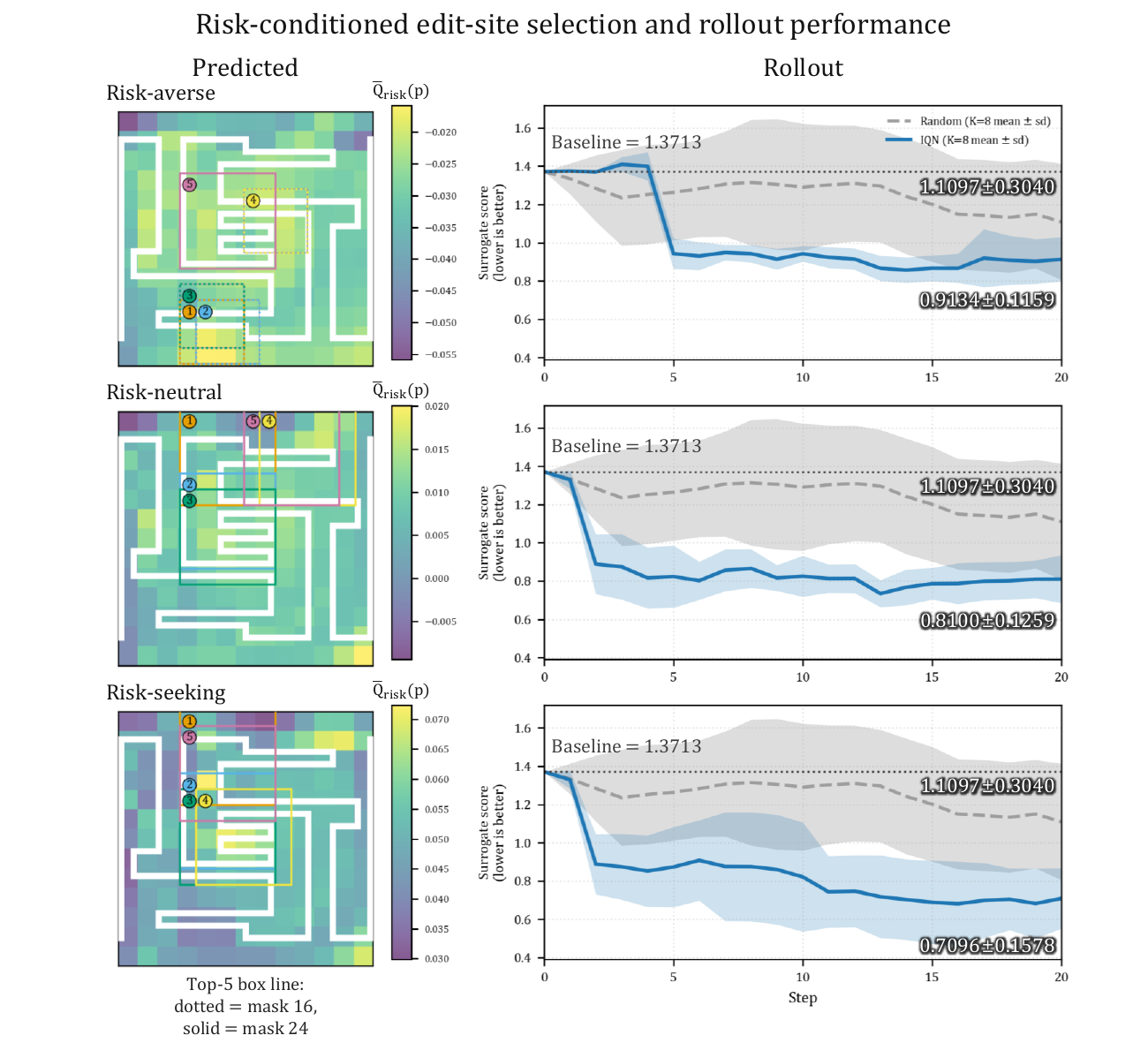}
\caption{Risk-conditioned edit-site selection and rollout performance on held-out design 7
(pair(16,24), IQN, $w{=}0.001$). Each row is one deployed risk band. Left: the position-value
surface $\bar Q_{\text{risk}}(p)$ of Eq.~\eqref{eq:policy} over the mask-16 candidate layer,
overlaid on the unedited skeleton, with the five positions the static policy visits first under
that band (numbered; dotted outline mask-16, solid outline mask-24). Right: surrogate score
against edit step, mean $\pm 1$ standard deviation over the same eight independent DDPM rollout
seeds as Fig.~\ref{fig:ssrcurve}, with random editing (grey) and the unedited-baseline score
(dotted line) as references; the three score panels share one $y$-axis. Step-20 means are
$0.913$ (averse), $0.810$ (neutral) and $0.710$ (seeking) against $1.110$ for random; every
IQN rollout improves on the baseline under all three bands, random in six of eight. The
seeking-band panel is the family solved with CFD in Table~\ref{tab:idx7_cfd} of the
supplementary material.}
\label{fig:riskrollout}
\end{figure}

Fig.~\ref{fig:riskrollout} gives the first single-design illustration, on the same held-out
design that \S\ref{sec:cfd_verify} solves with CFD. Moving the risk band from averse to seeking
changes which sites the static policy visits first, and the three rollouts reproduce, on one
design, the population-level ordering of Table~\ref{tab:seedspread}: the risk-seeking rollout
reaches the lowest step-20 score and the risk-averse rollout the highest, with the seed-to-seed
spread narrowing in the same direction. All three settings improve on the unedited baseline in
every rollout seed, against six of eight for random editing.

\subsubsection{CFD verification of the deployed policy}
\label{sec:cfd_verify}
\S\ref{sec:stage1} established surrogate fidelity on the labelled population. Stage 3 needs a
stronger claim, because the policy is evaluated on \emph{edited} designs, which
the surrogate never saw during training and which the policy actively steers toward the objective.
We therefore verified the surrogate against CFD on 240 policy-edited designs (four representative
conditions $\times$ 30 held-out images $\times$ \{agent, random\}).
\begin{itemize}
  \item \textbf{Decision-level agreement, edited designs.} Across the 240 CFD-verified designs,
        the surrogate and CFD agree on the \emph{sign} of agent-vs-random improvement in
        $116/120$ paired comparisons ($96.7\%$, $95\%$ CI $[91.7,98.7]$, $p{=}1.3\times10^{-29}$
        vs.\ chance); because these 120 comparisons reuse the same 30 held-out designs across four
        conditions rather than drawing on 120 independent designs, this interval assumes
        independence across conditions --- the same correlated-sample caveat noted for the
        mask-pooled estimate of \S\ref{sec:c1} --- and should be read descriptively. Magnitude
        agreement is Spearman $\rho{=}0.982$, Pearson $r{=}0.964$
        (within-image paired Pearson $r{=}0.962$). A Bland--Altman analysis shows the surrogate
        is systematically more optimistic for worse-scoring designs (proportional bias
        $r{=}-0.797$), which we flag as a limitation of the calibration, not the ranking.
  \item \textbf{Advantage is not exploited surrogate error.} Correlating the per-image policy
        advantage against four independent definitions of surrogate error, on the official
        $N{=}100$ held-out rollout across three masks, gives $|r|<0.11$ in all sixteen tests
        (Holm-adjusted $p\ge0.83$): the measured advantage is statistically unrelated to how
        wrong the surrogate is on that image, which argues against the concern that the policy
        is simply learning to exploit surrogate mis-scoring.
\end{itemize}
Both checks hold \emph{within the deployment range}: the surrogate's error grows
$3.2$--$3.4\times$ once the pressure-drop prediction is extrapolated outside the training
scaler's range, and both verification sets here stay inside that range by construction.

The two checks above are population-level, sampling one edited design per image. To confirm that
the surrogate scorer behind the seed-spread claims of \S\ref{sec:seedspread} tracks CFD seed by
seed, and not only design by design, we solved one $K{=}8$ rollout family in full --- an example
held-out design, the unedited baseline, and both arms at step 20 for all eight rollout seeds (17
designs; Fig.~\ref{fig:ssrcurve}, full detail in Table~\ref{tab:idx7_cfd} of the supplementary
material). The two
scorers agree closely (score mean absolute error $0.051$ across a range of $1.40$, $r{=}0.989$;
$0.03$~K in mean temperature, $3.7$~Pa in pressure drop) and put the agent ahead by a similar
margin on both (surrogate $+0.66$, CFD $+0.70$, against $+0.26$/$+0.20$ for random), agreeing on
the sign of the agent--random gap in seven of eight seeds and returning a seed-spread ratio near
two on both scorers ($1.93$ surrogate, $1.81$ CFD); with only eight seeds the $F(7,7)$ interval on
that ratio spans roughly $2.2\times$ either way and includes $1$, so no consistency claim rests on
this one design --- that is what Fig.~\ref{fig:seedspread_pct}, across the full held-out set, is
for. In physical units the agent lowers mean temperature by $0.26$~K and pressure drop by $9.6\%$
in all eight seeds, against $0.08$~K and $3.6\%$ for random, which \emph{raises} the corner count
instead of lowering it ($51\rightarrow55.6$ vs.\ $51\rightarrow46.3$); the temperature swing is
small in absolute terms but not on this problem's scale, since the labelled population spans only
about $0.6$~K in $T_{\text{avg}}$ (\S\ref{sec:stage1}), so this one edit covers roughly two-fifths
of that range, while uniformity is essentially unchanged ($0.7313\rightarrow0.7269$), consistent
with it being the quantity the agent trades away rather than improves. These are one design's
numbers, quoted to fix the scale of the effect rather than to estimate its size.

The same family also shows what the deterministic policy narrows and what it does not: all eight
rollouts share a bit-identical action sequence, yet end in eight distinct final layouts, because
the generator stays stochastic at every visited site --- the policy narrows the spread of outcome
\emph{quality}, not the diversity of designs, with the agent's eight CFD improvements clustering
at $+0.70$ (s.d.\ $0.18$) while random's spread ($0.33$) exceeds its own mean ($0.20$). Part of
that gap is structural rather than learned: the random arm re-draws positions at every seed and so
carries that variability on top of the generator's, while the agent's fixed visiting order does
not, so this one design cannot separate committing to a fixed order from committing to a good
one; that separation is available on the risk axis instead (\S\ref{sec:seedspread}), where moving
the dial from averse to seeking widens the agent's own seed-to-seed spread in all seven mask
configurations, so it is which order is fixed, not merely that one is, that moves consistency.

\begin{figure}[!htbp]\centering
\includegraphics[width=\textwidth]{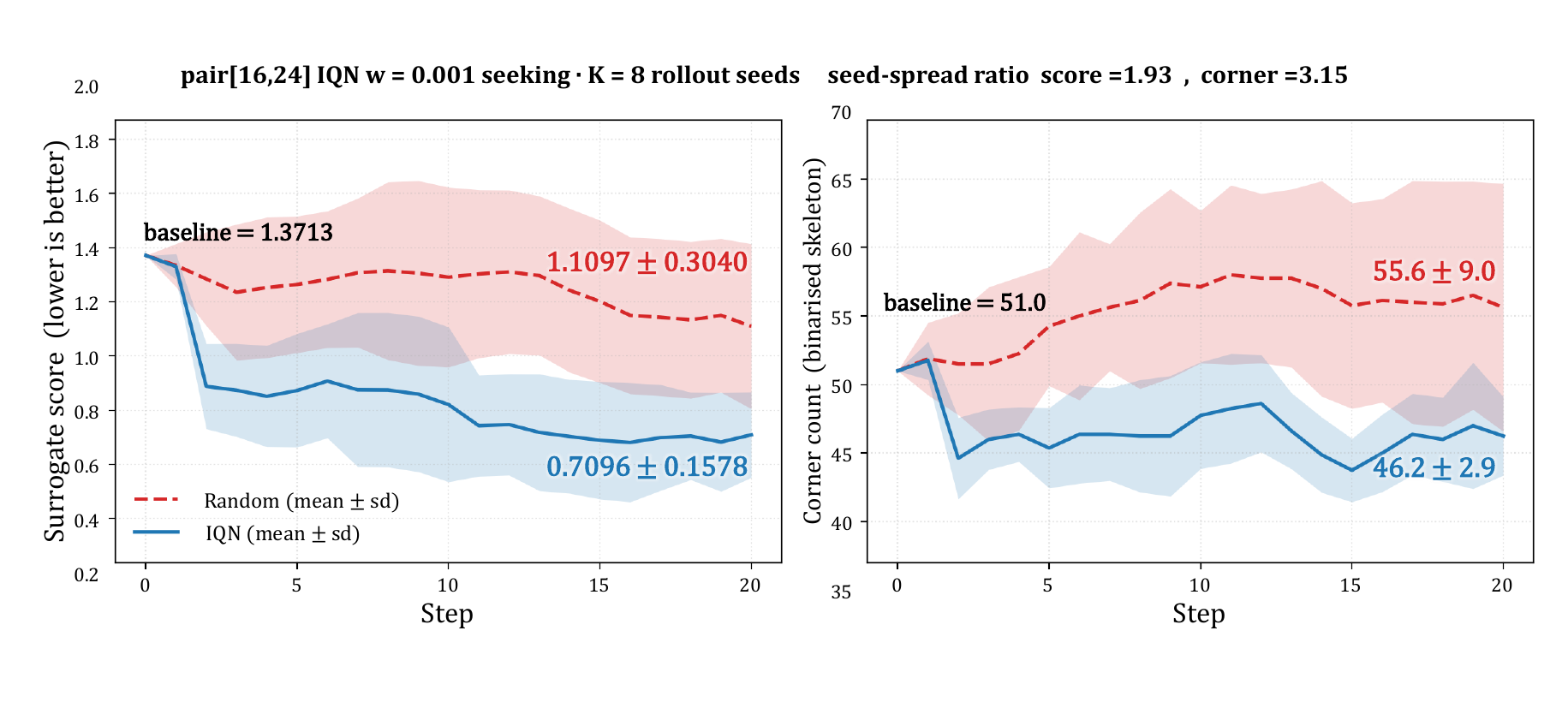}
\caption{The $K{=}8$ rollout family solved in full (held-out design 7, pair(16,24),
risk-seeking): surrogate score (left) and corner count (right) against edit step, mean $\pm1$
standard deviation over eight independent DDPM rollout seeds --- not a best-of-eight selection.
CFD was solved at step~0 and step~20 only, so it appears in Table~\ref{tab:idx7_cfd} of the
supplementary material rather than as a curve here.}
\label{fig:ssrcurve}
\end{figure}

\subsubsection{Supervised regression versus TD training}
\label{sec:tdvsup}
The design decision of \S\ref{sec:whynotrl} rests on a controlled comparison, holding the network
architecture and offline per-position outcome labels fixed and fitting the same quantile function
by (i) temporal-difference learning with a bootstrapped target, and (ii) supervised pinball-loss
regression, the rule used everywhere else in this paper. The two rank candidate positions equally
well (mean Spearman $\rho$ from $-0.012$ to $+0.088$ for the TD variants and $+0.145$ for the
supervised critic, all intervals overlapping), but supervised regression is markedly
better-calibrated along $\tau$: on a coarse five-point $\tau$ grid (the grid used by the TD
comparison harness), quantile-crossing falls from $9$--$32\%$ under TD to $1.0\%$ under
supervised training. Full tables, including a 100-point diagnostic $\tau$ grid on which all three
critics degrade to $41$--$48\%$ crossing, are given in the supplementary material
(\S\ref{app:crossing}).

\label{sec:calibration}
Even the supervised critic actually deployed, however, is not a calibrated probability
distribution: a probability-integral-transform (PIT) test rejects uniformity strongly across all
six significant conditions of Table~\ref{tab:c1} (Kolmogorov--Smirnov $p\approx0$; PIT mean
$0.36$--$0.41$ against an ideal of $0.5$),\footnote{Computed on the validation split of the
labelled training population, the only split with per-position outcome distributions; the
deployment evaluations of \S\ref{sec:c1} use the separate held-out set
(\S\ref{sec:data}).} and quantile-crossing within the deployed $\tau$ windows remains $7$--$20\%$.
Isotonic recalibration removes crossing only by construction and does not fix the underlying
miscalibration; it also significantly \emph{hurts} the ranking-based decision quality it is meant
to preserve ($p{=}0.00065$--$0.0024$). We therefore use the quantile function only as a ranking
device and a mean-improvement/consistency dial, not as a calibrated probability object.

\section{Discussion}
\label{sec:discussion}

\subsection{Interpretation of the results}
\label{sec:disc_why}

\paragraph{Ranking-based decisions versus calibrated absolute values}
A single pattern recurs across \S\ref{sec:results}: decisions that depend only on \emph{ordering}
(argsort the position-value map, rank agent vs.\ random by sign, use $\tau$ to move along a
frontier) are robust, while decisions that require a \emph{calibrated absolute value} (a
probability-integral-transform test, a coverage guarantee) are not. The comparison in
\S\ref{sec:tdvsup} is the clearest instance: for this task, where an offline pool of labelled edit
outcomes is cheap to build and a step's outcome does not depend on the trajectory that led to it,
sequential value-function training bought nothing on ranking and cost calibration, so a
distributional critic should be checked along $\tau$, not only across actions, since the metric
practitioners usually report (rank agreement with ground truth) is blind by construction to the
failure mode that matters for a risk-sensitive deployment. The same split shows up in the
per-position accuracy itself: the frozen linear term explains almost none of the per-position
variance (univariate correlations below $0.04$, Table~\ref{tab:features}), yet ranking positions
by it alone already wins against random editing in five of the seven mask configurations --- a
count that overstates the evidence, since those seven configurations yield only five distinct
ridge-only rollouts and no individual cell is significant (supplementary material,
\S\ref{app:arch}). Two measurements narrow that gap without closing it: the advantage is
front-loaded on the faster mask configurations, realised within the first five of 20 edits (the
horizon stays at 20 because the slowest configuration, pair(24,32), does not realise its advantage
until close to the full horizon); and on a held-out probe set for $m24$ the correlation between
predicted and realised position value is close to zero over all positions but roughly four times
larger once positions where no edit is realisable are excluded --- the model orders editable
positions weakly but does not identify which positions are editable at all. Both are narrowing
evidence, not a full mechanism, and come from a single mask configuration and a probe budget that
differs from deployment.

\begin{figure}[htbp]\centering
\includegraphics[width=0.85\textwidth]{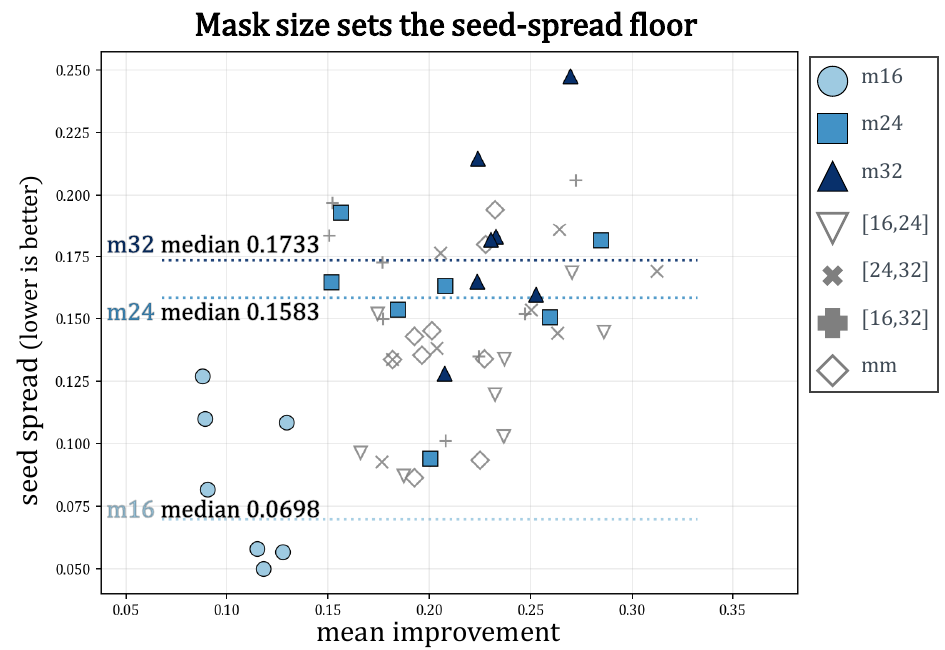}
\caption{The improvement--consistency plane, replotting the measurements of
\S\ref{sec:seedspread}. Each marker is one measured condition; dotted lines are the median seed
spread per single-mask configuration. Mask size sets a floor on achievable seed spread
($\rho{=}{+}0.833$ between mask size and the agent's seed spread, Table~\ref{tab:seedspread}), and
mean improvement rises with mask size as well ($\rho{=}{+}0.646$), so the two move together along
this axis. Improvement is the held-out mean at $N{=}100$; the spread is from the repeated-seed
campaign of \S\ref{sec:seedspread}.}
\label{fig:frontier_mask}
\end{figure}

\begin{figure}[htbp]\centering
\includegraphics[width=0.72\textwidth]{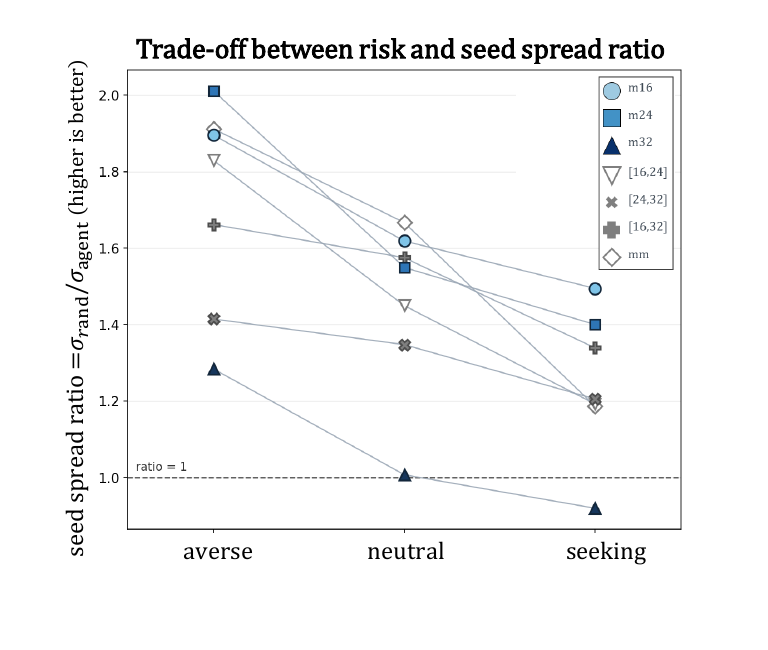}
\caption{The inference-time risk dial along the same frontier, replotting
\S\ref{sec:seedspread}. Each line is one mask configuration moved from risk-averse through
neutral to risk-seeking; the ratio falls in every configuration ($\rho{=}{-}0.564$,
Table~\ref{tab:seedspread}). The dashed line marks parity with the random baseline: $m32$ at
risk-seeking falls below it.}
\label{fig:frontier_risk}
\end{figure}

\paragraph{Risk dial $\tau$ versus corner-weight conditioning $w$}
This is why we present $\tau$ as a dial rather than as a claim that risk-seeking is the better
setting: \S\ref{sec:mechanism} and \S\ref{sec:seedspread} together show it moves the deployed
policy along a mean-improvement/consistency frontier, and which end of that frontier is preferable
is a deployment choice, not a result. The corner weight $w$ does not behave this way; it moves the
mean without moving the variance axis, so of the two inference-time conditioning inputs only
$\tau$ is a frontier dial. Mask size sets a \emph{floor} on that frontier that no risk setting
escapes: the median seed spread rises from $0.070$ ($m16$) to $0.158$ ($m24$) to $0.173$ ($m32$),
and larger masks buy their higher mean improvement by paying into that floor (mask size correlates
$+0.646$ with mean improvement and $+0.833$ with the spread across single-mask configurations),
making it the coarse control, chosen before deployment and not adjustable afterwards
(Fig.~\ref{fig:frontier_mask}). Within a fixed mask, moving $\tau$ from averse to seeking lowers
the seed-spread ratio monotonically in every configuration tested (Fig.~\ref{fig:frontier_risk}),
trading consistency for mean improvement along the frontier mask size positioned. A high
seed-spread ratio is not a general claim that a configuration is better, only that it is more
consistent than its own random baseline, and the ratio is not bounded below by one: the $m32$
configuration crosses below $\text{ratio}{=}1$ at risk-seeking, meaning the policy is \emph{less}
consistent across rollout seeds than random editing there even though it retains a mean-improvement
advantage --- exactly the combination a single-number report of mean improvement would hide.


\paragraph{Single-stroke connectivity and the realised-edit rate}
Because a step only changes the layout when a sampled candidate satisfies the single-stroke
constraint (\S\ref{sec:protocol}), part of the edit budget produces no change at all. This is one
reason every position is scored and ranked rather than sampled: ranking consumes no additional
generative calls, so the budget can be spent where the model rates the prospects highest instead
of on arbitrary positions. We do not attach a number to the realised-edit rate here: it is not
logged uniformly across the campaign, so any single figure we could quote would rest on a
different measurement basis than the headline results, and pairing the two would invite exactly
the mismatched comparison this paper is otherwise careful to avoid. The single-stroke constraint
does not make the search space too restrictive to improve in: under the same constraint,
unguided best-of-8 search lowers the surrogate score from $1.417$ to $0.131$ over 200
generations (\S\ref{sec:stage2}), while CFD confirms the direction for $m24$ and $m32$. This
indicates that substantial improvement remains possible within the constrained design space.

\subsection{Limitations}
\label{sec:disc_limits}

\paragraph{Surrogate extrapolation}
The CFD checks of \S\ref{sec:cfd_verify} license the surrogate as a \emph{decision} proxy only
within the pressure-drop range it was trained on, where its error is small and its sign agreement
with CFD is high. Outside that range the error grows several-fold, so staying in-range is an
explicit deployment gate rather than an incidental property of these particular test sets.

\paragraph{Scope: operating point, seed coverage, and architecture comparison}
This study covers a single physical operating point (fixed heat flux). A corner-weight
conditioning sweep is confounded with a scoring-objective mismatch (supplementary
\S\ref{app:cornerweight}), and
the quantile function itself is not calibrated (\S\ref{sec:calibration}). The headline effect size
depends measurably on which of several plausible rollout seeds is used (\S\ref{sec:rolloutseed}),
where the pre-registered seed is the largest of the five measured. For the DQN and ridge-only
arms, the architecture comparison still rests on a single training seed --- the IQN-vs-FQF
comparison is now three-seed at the matched cell (supplementary material, \S\ref{app:arch}), but
DQN and ridge-only are not. This study also compares only against random editing; a published
external baseline \citep{kimhan2023} was judged out of scope for the reasons given in
\S\ref{sec:rw_cooling}.

\subsection{Implications for practice}
\label{sec:disc_implications}

\paragraph{For generative-design agents} Practitioners deploying a similar system should restrict
themselves to ranking-based uses of a learned distributional value function unless they have
separately verified calibration; any use requiring calibrated absolute probabilities (e.g.\
$P(\text{improvement}>t)$, or an adaptive stopping rule) should not be attempted on top of a
function like ours without first solving the calibration problem, which we did not solve.

\paragraph{For offline outcome pools} The same task structure that made supervised regression
preferable here --- a cheap-to-build offline pool of labelled outcomes, and a step's outcome that
does not depend on the trajectory that led to it --- recurs in other generative-design editing
problems; where it holds, sequential training is a cost to justify, not a default.

\paragraph{For surrogate verification} A deployment that begins to produce designs outside a
surrogate's training range should re-verify against the ground-truth solver before trusting its
rankings, rather than assuming in-range fidelity extends to out-of-range decisions.

\section{Conclusion}
\label{sec:conclusion}
We presented a cooling-channel design-support system that uses the query interface of a
distributional value function, IQN's $\tau$-conditioned implicit quantile network, without
its usual reinforcement-learning training loop, fitting a frozen linear term plus a residual by
supervised regression on offline edit outcomes and deploying it as a deterministic, one-shot
ranking policy. On a held-out, paired, 100-image evaluation the policy beats random editing in six
of seven mask configurations, with a headline effect of $+0.1389$
($p{=}7.8\times10^{-11}$) at the pre-registered protocol, an effect size that falls to $+0.094$
when averaged over five rollout seeds; a single inference-time risk dial moves the policy along a mean-
improvement/consistency frontier that architecture identity does not touch; and the surrogate
objective that drives training is CFD-verified to be a valid decision proxy within its deployment
range. The failures are reported with the same care as the successes: temporal-difference training
measurably hurts distributional calibration relative to supervised regression on this task, the
learned quantile function fails a calibration test regardless of training method, and
conditioning the policy on a value that does not match the true scoring objective costs
performance. The proposed framework improves design quality under a constrained
single-stroke topology while explicitly exposing the trade-off between improvement and
consistency. Beyond this application, we
conjecture that the same recipe --- fitting a value function by
supervised regression when an offline outcome pool is available, using it only for the decisions
that survive miscalibration, and verifying a learned surrogate against physics before trusting a
policy trained on it --- extends to other topology-constrained generative-design problems that
share the two conditions of \S\ref{sec:disc_implications}: a cheap-to-build offline pool of
labelled outcomes, and a step's outcome that does not depend on the trajectory that produced it.

\section*{Declaration of competing interest}
\label{sec:coi}
The authors declare that they have no known competing financial interests or personal
relationships that could have appeared to influence the work reported in this paper.

\section*{Declaration of generative AI use}
\label{sec:aiuse}
During the preparation of this work the authors used a generative AI tool to translate
the authors' native-language drafting into English. After using this tool, the authors reviewed
and edited the content as needed and take full responsibility for the content of the published
article.

\section*{Acknowledgments}
\label{sec:ack}
This work was supported by grants from the Ministry of Science and ICT (GTL24033-000, N10250154,
No.~2022-0-00986, and RS-2026-25537227), the Ministry of Trade, Industry and Energy
(RS-2025-02317327 and RS-2025-25444634), the Ministry of Oceans and Fisheries (PET0050), and
Korea Hydro \& Nuclear Power Co., Ltd.\ (No.~8-Tech-07).

\section*{CRediT authorship contribution statement}
\label{sec:credit}
\begin{itemize}
  \item \textbf{Leekyo Jeong:} Conceptualization, Methodology, Software, Validation, Formal
        analysis, Investigation, Data curation, Writing -- original draft, Writing -- review \&
        editing, Visualization.
  \item \textbf{Yoon Koo Lee:} Investigation, Validation, Writing -- review \& editing.
  \item \textbf{Namwoo Kang:} Supervision, Funding acquisition, Writing -- review \& editing.
\end{itemize}

\section*{Data availability}
\label{sec:repro}
The data that support the findings of this study are not publicly available. They are available
from the corresponding author upon reasonable request.


\end{document}